\documentclass[sigconf]{acmart}

\usepackage{soul}
\usepackage{url}
\usepackage{hyperref}
\usepackage{graphicx}
\usepackage{amsmath,amsthm}
\usepackage{color}
\usepackage{mathtools}
\usepackage{booktabs}
\usepackage[caption=false]{subfig}
\usepackage{multirow}

\newcommand{\ignore}[1]{}
\newcommand{\specialcell}[2][c]{\begin{tabular}[#1]{@{}c@{}}#2\end{tabular}}
\newcommand{\vthead}[1]{\makebox[1em][l]{\rotatebox{90}{#1}}}

\newtheorem{insight}{Insight}
\newtheorem{definition}{Definition}

\DeclareMathOperator{\sigmoid}{sigmoid}

\DeclareMathOperator{\bern}{Bernoulli}

\DeclareMathOperator{\NLL}{NLL}
\DeclareMathOperator{\brier}{Brier}
\DeclareMathOperator{\kl}{KL}

\AtBeginDocument{%
  \providecommand\BibTeX{{%
    \normalfont B\kern-0.5em{\scshape i\kern-0.25em b}\kern-0.8em\TeX}}}

\copyrightyear{2021}
\acmYear{2021}
\setcopyright{acmcopyright}\acmConference[ACSAC '21]{Annual Computer Security Applications Conference}{December 6--10, 2021}{Virtual Event, USA}
\acmBooktitle{Annual Computer Security Applications Conference (ACSAC '21), December 6--10, 2021, Virtual Event, USA}
\acmPrice{15.00}
\acmDOI{10.1145/3485832.3485916}
\acmISBN{978-1-4503-8579-4/21/12}

\begin{document}

\title{Can We Leverage Predictive Uncertainty to Detect Dataset Shift and Adversarial Examples in Android Malware Detection?}
\author{Deqiang Li}
\affiliation{%
  \institution{Nanjing University of Science and Technology}
  \city{Nanjing}
  \country{China}
}
\email{lideqiang@njust.edu.cn}

\author{Tian Qiu}
\affiliation{%
  \institution{Nanjing University of Science and Technology}
  \city{Nanjing}
  \country{China}
}
\email{qiutian@njust.edu.cn}

\author{Shuo Chen}
\affiliation{%
  \institution{RIKEN}
  \city{Saitama}
  \country{Japan}
}
\email{shuo.chen.ya@riken.jp}

\author{Qianmu Li}
\affiliation{%
	\institution{Nanjing University of Science and Technology}
	\city{Nanjing}
	\country{China}
}
\email{qianmu@njust.edu.cn}
\additionalaffiliation{%
	\institution{Wuyi University}
	\city{Jiangmen}
	\country{China}
}

\author{Shouhuai Xu}
\affiliation{%
  \institution{University of Colorado Colorado Springs}
  \streetaddress{1420 Austin Bluffs Pkwy}
  \city{Colorado Springs}
  \state{Colorado}
  \country{USA}
}
\email{sxu@uccs.edu}

\renewcommand{\shortauthors}{D. Li, T. Qiu, S. Chen, Q. Li, and S. Xu}

\begin{abstract}
The deep learning approach to detecting {\em mal}icious soft{\em ware} (malware) is promising but has yet to tackle the problem of {\em dataset shift}, namely that the joint distribution of examples and their labels associated with the test set is different from that of the training set. This problem causes the degradation of deep learning models {\em without} users' notice. In order to alleviate the problem, one approach is to let a classifier not only predict the label on a given example but also present its uncertainty (or confidence) on the predicted label, whereby a defender can decide whether to use the predicted label or not. While intuitive and clearly important, the capabilities and limitations of this approach have not been well understood. In this paper, we conduct an empirical study to evaluate the quality of predictive uncertainties of malware detectors. Specifically, we re-design and build 24 Android malware detectors (by transforming four off-the-shelf detectors with six calibration methods) and quantify their uncertainties with nine metrics, including three metrics dealing with data imbalance. Our main findings are: (i) predictive uncertainty indeed helps achieve reliable malware detection in the presence of dataset shift, but cannot cope with adversarial evasion attacks; (ii) approximate Bayesian methods are promising to calibrate and generalize malware detectors to deal with dataset shift, but cannot cope with adversarial evasion attacks; (iii) adversarial evasion attacks can render calibration methods useless, and it is an open problem to quantify the uncertainty associated with the predicted labels of adversarial examples (i.e., it is not effective to use predictive uncertainty to detect adversarial examples).
\end{abstract}



\maketitle

\section{Introduction}

Malware is a big threat to cybersecurity. Despite tremendous efforts, communities still suffer from this problem because the number of malware examples consistently increases. For example, Kaspersky \cite{kaspersky:Online} detected 21.6 million unique malware in 2018, 24.6 million in 2019, and 33.4 million in 2020. This highlights the necessity of automating malware detection via machine learning techniques \cite{DBLP:journals/csur/YeLAI17}. 
However, a fundamental assumption made by machine learning is that the training data distribution and the test data distribution are identical. In practice, this assumption is often invalid because of the {\em dataset shift} problem \cite{10.5555/1462129}.
Note that dataset shift is related to, but different from, the {\em concept drift} problem, which emphasizes that the conditional distribution of the labels conditioned on the input associated with the test data is different from its counterpart associated with the training data \cite{DBLP:journals/ml/WidmerK96}.
While some people use these two terms interchangeably \cite{zliobaite2016,DBLP:journals/tkde/LuLDGGZ19}, we 
choose to use dataset shift because it is a broader concept than concept drift.

Several approaches have been proposed for alleviating the dataset shift problem, such as: periodically retraining malware detectors \cite{DBLP:conf/cns/XuZXY14,DBLP:conf/eisic/ChenYB17}, extracting invariant features \cite{10.1145-3372297.3417291}, and detecting example anomalies \cite{203684}.
One fundamental open problem is to quantify the {\em uncertainty} that is inherent to the outcomes of malware detectors (or the confidence a detector has in its prediction).
A well-calibrated uncertainty indicates the potential risk of the accuracy decrease and enables malware analysts to conduct informative decisions (i.e., using the predicted label or not). One may argue that a deep learning model associates its label space with a probability distribution. However, this ``as is'' method is poorly calibrated and causes poor uncertainty estimates \cite{guo2017calibration}. This problem has motivated researchers to propose multiple methods to calibrate the probabilities, such as: variational Bayesian inference \cite{blundell2015weight}, Monte Carlo dropout \cite{gal2016dropout}, stochastic gradient MCMC \cite{welling2011bayesian}, and non-Bayesian methods such as ensemble \cite{lakshminarayanan2017simple}. Quantifying the uncertainty associated with predicted labels (i.e., predictive uncertainty) in the presence of dataset shift has received a due amount of attention in the context of image classification \cite{blundell2015weight,lakshminarayanan2017simple}, medical diagnoses \cite{leibig2017leveraging}, and natural language processing \cite{snoek2019can}, but not in the context of malware detection. This motivates us to answer the question in the title.

\smallskip
\noindent{\bf Our contributions}. In this paper, we empirically quantify the predictive uncertainty of 24 deep learning-based Android malware detectors.
These detectors correspond to combinations of four malware detectors, which include two {\em multiple layer perceptron} based methods (i.e., DeepDrebin \cite{grosse2017adversarial} and MultimodalNN \cite{8443370}) and one {\em convolutional neural network} based method (DeepDroid \cite{10.1145/3029806.3029823}), and the {\em recurrent neural network} based method (Droidetec \cite{ma2020droidetec}); 
and six {\em calibration} methods, which are vanilla (no effort made for calibration), {\em temperature scaling} \cite{guo2017calibration}, Monte Carlo (MC) dropout \cite{gal2016dropout}, Variational Bayesian Inference (VBI) \cite{snoek2019can},
deep ensemble \cite{lakshminarayanan2017simple} and its weighted version. The vanilla and temperature scaling calibration methods belong to the post-hoc strategy, while the others are ad-hoc and require us to transform the layers of an original neural network in a principled manner (e.g., sampling parameters from a learned distribution). In order to evaluate the quality of predictive uncertainty on {\em imbalanced} datasets, we propose useful variants of three standard metrics.

By applying the aforementioned 24 malware detectors to three Android malware datasets, we draw the following insights: 
(i) We can leverage predictive uncertainty to achieve reliable malware detection in the presence of dataset shift to some extent,
while noting that a defender should trust the predicted labels with uncertainty below a certain threshold.
(ii) Approximate Bayesian methods are promising to calibrate and generalize malware detectors to deal with dataset shift.
(iii) Adversarial evasion attacks can render calibration methods useless and thus the predictive uncertainty.

We have made the code of our framework publicly available at \url{https://github.com/deqangss/malware-uncertainty}.
It is worth mentioning that after the present paper is accepted, we became aware of a preprint  \cite{DBLP:journals/corr/abs-2108-04081}, which investigates how to leverage predictive uncertainty to deal with false positives of Windows malware detectors, rather than dealing with dataset shift issues.

\section{Problem Statement} \label{sec:problem-setup}

\subsection{Android Apps and Malware Detection}

Since Android malware is a major problem and deep learning is a promising technique, our empirical study focuses on Android malware detection. Recall that an Android Package Kit (APK) is a zipped file that mainly contains:
(i) {\em AndroidManifest.xml}, which declares various kinds of attributes about the APK (e.g., {\em package} name, required {\em permissions}, {\em activities}, {\em services}, {\em hardware});
(ii) {\em classes.dex}, which contains the APK's functionalities that can be understood by the Java virtual machines (e.g., Android Runtime); 
(iii) {\em res} folder and {resources.arsc}, which contain the resources used by an APK;
(iv) {\em lib} folder, which contains the native binaries compatible to different Central Processing Unit (CPU) architectures (e.g., ARM and x86);
(v) {\em META-INF} folder, which contains the signatures of an APK.
For analysis purposes, one can unzip an APK to obtain its files in the binary format, or disassemble an APK by using an appropriate tool (e.g., Apktool \cite{apktool:Online} and Androguard \cite{Androguard:Online}), to obtain human-readable codes and manifest data (e.g., {\em AndroidManifest.xml}).

A malware detector is often modeled as a supervised binary classifier that labels an example as benign (`0') or malicious (`1'). Let $\mathcal{Z}$ denote the example space (i.e., consisting of benign and malicious software) and $\mathcal{Y}=\{0,1\}$ denote the label space. Let $p^\ast(z,y)=p^\ast(y|z)p^\ast(z)$ denote the underlying joint distribution, where $z\in \mathcal{Z}$ and $y\in \mathcal{Y}$. The task of malware detection deals with the conditional distribution $p^\ast(y|z)$. Specifically, given a software sample $z\in\mathcal{Z}$, we consider Deep Neural Network (DNN) $p(y=1|z,\theta)$, which uses the {\em sigmoid} function in its output layer, to model $p^\ast(y|z)$, where $\theta$ represents the learnable parameters. A detector denoted by $f:\mathcal{Z}\to\mathcal{Y}$, which consists of DNNs and is learned from the training set $D_{train}$,  returns 1 if $p(y=1|z,\theta)\geq 0.5$ and 0 otherwise. We obtain $p(y=0|z)=1-p(y=1|z,\theta)$.
Moreover, let $p'(z,y)$ denote the underlying distribution of test dataset $D_{test}$. 

\subsection{Problem Statement}

There are two kinds of uncertainty associated with machine learning: {\em epistemic} vs. {\em aleatoric} \cite{snoek2019can}.
The {\em epistemic} uncertainty tells us about which region of the input space is not perceived by a model \cite{DBLP:journals/corr/abs-2011-07586}. That is, a data sample in the dense region will get a low epistemic uncertainty and get a high epistemic uncertainty in the sparse region. On the other hand, the {\em aleatoric} uncertainty is triggered by data noises and is not investigated in this paper.

It is widely assumed that $p'(z,y)=p^*(z,y)$ in malware detection and in the broader context of machine learning. However, this assumption does not hold in the presence of {\em dataset shift}, which reflects non-stationary environments (e.g., data distribution changed over time or the presence of adversarial evasion attacks) \cite{10.5555/1462129}.  Most dataset shifts possibly incur changes in terms of epistemic uncertainty (excluding label flipping or concept of input $z$ changed thoroughly). We consider three settings that potentially trigger the $p'(z,y)\neq p^\ast(z,y)$:
\begin{itemize}
	\item {\em Out of source}: The $D_{test}$ and $D_{train}$ are drawn from different sources \cite{demontis2017yes}. 

	\item {\em Temporal covariate  shift}: The test data or $p'(z)$ evolves over time \cite{235493}.
	
	\item {\em Adversarial evasion attack}: The test data is manipulated adversarially; i.e.,  $(z',y=1)\sim p'(z,y=1)$ is perturbed from $(z,y=1)\sim p^\ast(z,y=1)$, where $z'$ and $z$ have the same functionality and `$\sim$' denotes ``is sampled from'' \cite{li2020adversarial}.
\end{itemize}
One approach to coping with these kinds of dataset shifts is to make a malware detector additionally quantify the uncertainty associated with a prediction so that a decision-maker decides whether to use the prediction \cite{lakshminarayanan2017simple,snoek2019can,DBLP:conf/iclr/WangLN21}. 
This means that a detector $p(y|z,\theta)$ should be able to model $p^\ast(y|z)$ well while predicting examples $z$ of $(z,y)\sim p'(z,y)$ with high  uncertainties when $(z,y) \not\sim p^\ast(z,y)$, where $\not \sim$ denotes ``does not obey the distribution of''. Specifically, the quality of prediction uncertainty can be assessed by their confidence scores. This can be achieved by leveraging the notion of {\em calibration} \cite{lakshminarayanan2017simple}. A malware detector is said to be {\em well-calibrated} if the detector returns a same confidence score $q\in[0,1]$ for a set of test examples $Z_q\subseteq \mathcal{Z}$, in which the malware examples are distributed as $q$
\cite{vaicenavicius2019evaluating}. Formally, let $\Pr(y=1|Z_q)$ denote the proportion of malware examples in the set $Z_q$ that are indeed malicious. 
We have
\begin{definition}[Well-calibrated malware detector, adapted from \cite{vaicenavicius2019evaluating}]
	\label{def:well-calibration}
A probabilistic malware detector $p(\cdot,\theta):\mathcal{Z}\to[0,1]$ is well-calibrated if for each confidence $q\in[0,1]$ and
$Z_q=\{z:p(y=1|z, \theta)=q, \forall z\in\mathcal{Z}\}$, it holds that $\Pr(y=1|Z_q)=q$.
\end{definition}
Note that Definition \ref{def:well-calibration} means that for a well-calibrated detector, the fraction of malware examples in example set $Z_q$ is indeed $q$. This means that we can use $q$ as a confidence score for quantifying uncertainty when assessing the trustworthiness of a detector's predictions. This also implies that {\em detection accuracy} and {\em well-calibration} are two different concepts. This is so because an accurate detector is not necessarily well-calibrated (e.g., when correct predictions with confidence scores near 0.5). Moreover, a well-calibrated malware detector is also not necessarily accurate.

It would be ideal if we can rigorously quantify or prove the uncertainty (or bounds) associated with a malware detector. However, this turns out to be a big challenge that has yet to be tackled. This motivates us to conduct an empirical study to characterize the uncertainty associated with Android malware detectors. Hopefully the empirical findings will contribute to theoretical breakthrough in the near future.

Specifically, our empirical study is centered at the question in title, which is further disintegrated as four Research Questions (RQs) as follows:

\begin{itemize}
	\item{\bf RQ1}: What is the predictive uncertainty of malware detectors in the absence of dataset shift?
	
	\item{\bf RQ2}: What is the predictive uncertainty of malware detectors with respect to out-of-source examples?
	
	\item{\bf RQ3}: What is the predictive uncertainty of malware detectors under temporal covariate  shift?
	
	\item{\bf RQ4}: What is the predictive uncertainty of malware detectors under adversarial evasion attacks?
\end{itemize}
Towards answering these questions, we need to empirically study the predictive distribution $p(y|z,\theta)$ of 
malware detectors in a number of scenarios.

\section{Empirical Analysis Methodology} \label{sec:methodology}

In order to design a competent methodology, we need to select malware detector that are accurate in detecting malware, select the calibration methods that are appropriate for these detectors, and metrics. This is important because a ``blindly'' designed methodology would suffer from incompatibility issues (e.g., integrating different feature extractions modularly, or integrating post- and ad-hoc calibration methods together). The methodology will be designed in a modular way so that it can be extended to accommodate other models or calibration methods in a plug-and-play fashion.

\subsection{Selecting Candidate Detectors}

We focus on deep learning based Android malware detectors and more specifically choose the following four Android malware detectors.

\noindent
\textbf{DeepDrebin} \cite{grosse2017adversarial}: It is a Multiple Layer Perceptron (MLP)-based malware detector learned from the Drebin features \cite{arp2014drebin}, which are extracted from the {\em AndroidManifest.xml} file and the {\em classes.dex} file reviewed above. These features are represented by binary vectors, with each element indicating the presence or absence of a feature.

\noindent
\textbf{MultimodalNN} \cite{8443370}: It is a multimodal detector that contains five headers and an integrated part, all of which are realized by MLP. The 5 headers respectively learn from 5 kinds of features: (i) {\em permission-component-environment} features extracted from the manifest file; (ii) {\em strings} (e.g., IP address); (iii) system APIs; (iv) {\em Dalvik} opcode; and (v) ARM opcodes from native binaries. These features are represented by their occurrence frequencies. These 5 headers produce 5 pieces of high-level representations, which are concatenated to pass through the integrated part for classification.

\noindent
\textbf{DeepDroid} \cite{10.1145/3029806.3029823}: It is Convolutional Neural Network (CNN)-based and uses the TextCNN architecture \cite{kim2014convolutional}. The features are {\em Dalvik} opcode sequences of {\em smali} codes. 

\noindent
\textbf{Droidetec} \cite{ma2020droidetec}  is an RNN-based malware detector with the architecture of Bi-directional Long Short Term Memory (Bi-LSTM) \cite{thireou2007bidirectional}. Droidetec partitions a Function Call Graph (FCG) into sequences according to the caller-callee relation and then concatenates these sequences for passing through the Bi-LSTM model.

These detectors leverage several types of deep learning models. Indeed, we also implement an end-to-end method ({\em R2-D2} \cite{8622324}) and find it ineffectiveness due to the over-fitting issue. Therefore, we eliminate it from our study.

\ignore{
	\item {\em R2-D2} \cite{8622324} is a also convolutional neural network based malware detector. Compared to DeepDroid, this method learns from RGB images (three dimensional matrices with value in the range of $[0,255]$) which is mapped from the dex files of an app in binary format. After comparing with several CNNs, GoogleNet is selected to implement the malware detectors.
}

\subsection{Selecting Calibration Methods}

In order to select calibration methods to calibrate the malware detectors, the following two criteria can be used: (i) they are known to be effective and (ii) they are scalable for learning a deep learning model.
In particular, the preceding highlights the importance of the computational complexity of a calibration method. Based on a previous study \cite{snoek2019can}, these criteria lead us to select the following six methods.

\noindent
\textbf{Vanilla}: It serves as a baseline and means that no effort is made to calibrate a model $p(y=1|z, \theta)$. 

\noindent
\textbf{Temperature scaling} (Temp scaling) \cite{guo2017calibration}: It is a post-processing method that learns extra parameters to scale the {\em logits} of deep learning models on the validation set, where {\em logits} are the input of the activation  of $\sigmoid$. 

\noindent
\textbf{Monte Carlo dropout} (MC dropout) \cite{gal2016dropout}: It technically adds a {\em dropout} layer \cite{srivastava2014dropout} before the input of every layer contained in a model. This dropout operation is performed in both the training and test phases different from the normal usage that turns dropout on in the training phase and off in the test phase. In theory, MC dropout is an approximate Bayesian inference, which takes as input an example $z$ and marginalizes the parameters $\theta$ out for returning the predictive probability:
\begin{align}
	p(y=1|z,D_{train})=\int~&p(y=1|z,\theta)p(\theta|D_{train})d\theta \label{eq:bay-ens}
\end{align}
Due to the intricate neural networks, an analytical solution to obtaining $p(\theta|D_{train})$ is absent. One alternative is to approximate $p(\theta|D_{train})$ via a known functional form distribution $q(\omega)$ with variables $\omega$, leading to	
$p(y|z,D_{train})\approx\mathbb{E}_{q(\theta|\omega)}p(y|z,\theta)$. Specifically, let $\theta=\{\mathbf{W}_i\}_{i=1}^{l}$ be the set of $l$ parameters of a neural network, MC dropout defines $q(\omega)$ as:
\begin{eqnarray}
    \mathbf{W}_i = \mathbf{M}_i \cdot \mathbf{V}_i ~~\text{with}~~\mathbf{V}_i\sim \bern(r_i) ,
\end{eqnarray}
where $\mathbf{M}_i$ is learnable variables, entities of $\mathbf{V}_i$ are sampled from a Bernoulli distribution with the probability $r_i$ (i.e., dropout rate), and $\omega=\{\mathbf{M}_i,r_i\}_{i=1}^l$. In the training phase, {\em variational learning} is leveraged to look for $\{\mathbf{M}_i\}_{i=1}^l$ \cite{DBLP:conf/nips/Graves11,blundell2015weight}, which minimizes the Kullback-Leibler (KL) divergence between $p(\theta|D_{train})$ and $q(\omega)$. In the test phase, Eq.\eqref{eq:bay-ens} is degraded by averaging $T(T>0)$ models $\{\theta^j\}_{j=1}^T$ (each of which is sampled from $q(\omega)$), namely:
\begin{equation}
	p(y=1|z)=\frac{1}{T}\sum_{j=1}^T~p(y=1|z, \theta^i). \label{map-pred}
\end{equation}
Eq.\ref{map-pred} says $p(y=1|z)$ is obtained just by keeping the dropout switched and averaging the results of $T$ times predictions. 

\noindent
\textbf{Variational Bayesian Inference} (VBI) \cite{snoek2019can}: It is also an approximate Bayesian method. Distinguishing from MC dropout, the parameters $\theta$ of neural network are directly sampled from a known form distribution $q(\omega)$ with variables $\omega$. That is 
\begin{eqnarray}
	\mathbf{W}\sim q(\omega)~~\text{with}~~ \mathbf{W}\in\theta
\end{eqnarray} 
The training and test manner is the same as MC dropout.

\noindent
\textbf{Deep Ensemble} \cite{lakshminarayanan2017simple}: It learns $T$ independent neural networks, which are diversified by randomly initialized parameters. The intuition behind this idea is that {\em dropout} itself is an ensemble \cite{srivastava2014dropout}. 
 
\noindent
\textbf{Weighted Deep Ensemble} (wEnsemble): It has the same setting as Deep Ensemble, except for the weighted voting 
\begin{eqnarray}
    p(y|z)=\sum_{i=1}^T~w_i p(y=1|z, \theta^i) \label{eq:wens}
\end{eqnarray}
with $w_i \geq 0$ and $\sum_{i=1}^T{w_i}=1$. 

\subsection{Calibrating Detectors}

\begin{figure}[!htbp]
	\centering
	\scalebox{0.52}{
		\includegraphics{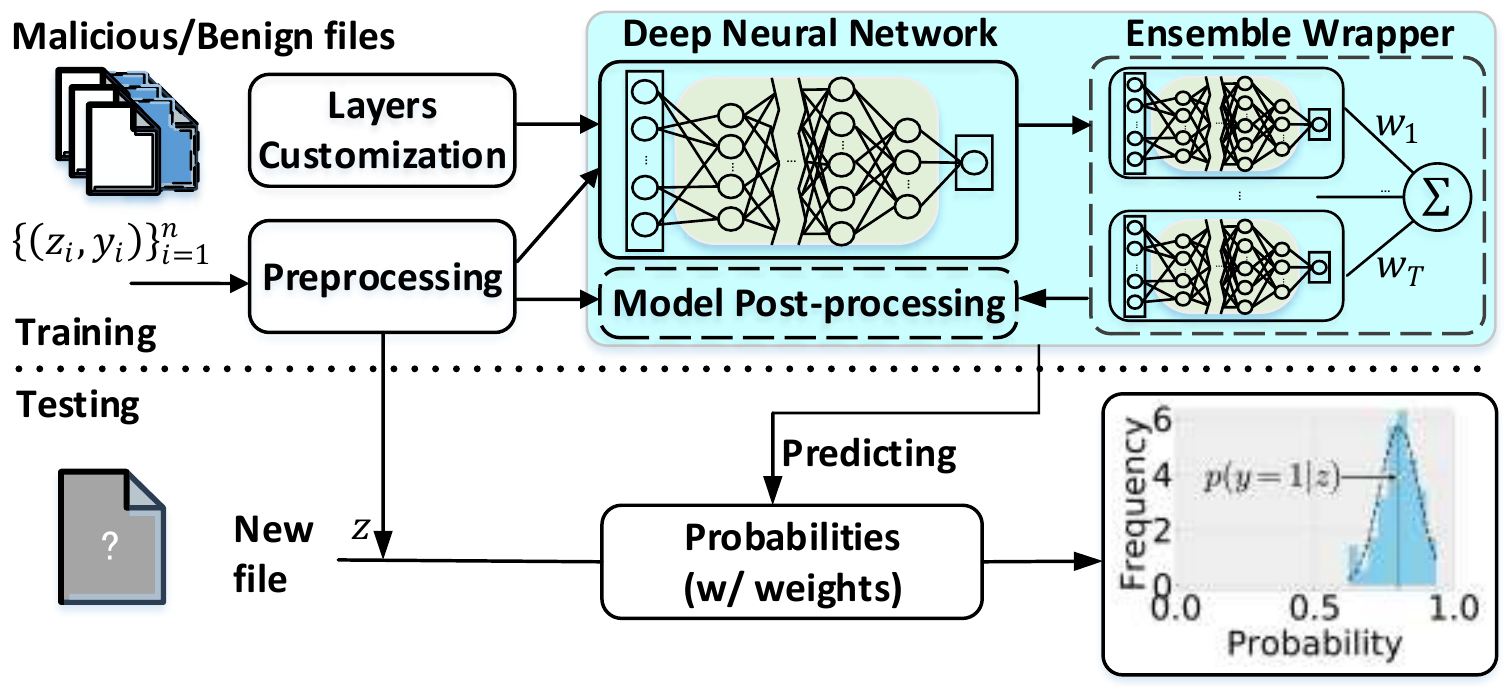}
	}
	\caption{Our methodology for calibrating deep malware detectors to answer the aforementioned RQ1-RQ4, where dashed boxes indicate that the modules may not be necessary for some calibration methods. In the training phase, we learn the calibrated malware detectors. In the test phase, predicted probabilities (with weights if applicable) are obtained.
	}
	\label{fig:ensemble}
\end{figure}

Figure \ref{fig:ensemble} highlights our methodology in calibrating deep malware detectors for answering the research questions mentioned above (i.e., RQ1-RQ4). Each calibration method is calibrated into each of the select detectors. The methodology has a training phase and a testing phase. Each phase has five modules: {\em preprocessing}, {\em layers customization}, {\em deep neural network}, {\em ensemble wrapper}, and {\em model post-processing}, which are described below.

The {\em preprocessing} module transforms program files into the data formats that can be processed by deep learning models, including feature extraction and low-level feature representation.
The {\em layers customization} module modifies the layers of standard deep learning models to incorporate appropriate calibration methods, such as placing a {\em dropout} layer before the input of the fully connected layer, the convolutional layer, or the LSTM layer; sampling parameters from learnable distributions. 
The {\em deep neural network} module constructs the deep learning models with the customized layers mentioned above to process the preprocessed data for training or testing purposes.
The {\em ensemble wrapper} module uses an ensemble of deep learning models in the training and testing phases according to the incorporated calibration method. In this module, we load basic block models sequentially in the training and test phases for relieving the memory complexity. 
The {\em model post-processing} module copes with the requirements of post-hoc calibration methods. The input to this module is the detector's output (i.e., predicted probability that a file belongs to which class) and a validation dataset (which is a fraction of data sampled from the same distribution as the training set). This means that this module does not affect the parameters $\theta$ of neural networks. 

\subsection{Selecting Metrics} \label{sec:metrics}

In order to quantify the predictive uncertainty of a calibrated detector, we need to use a test dataset, denoted by $D_{test}$, and some metrics. In order to make the methodology widely applicable, we consider two scenarios: the ground-truth labels of testing dataset are available vs. are not available. 
It is important to make this distinction is important because ground-truth labels, when available, can be used to validate the trustworthiness of detectors' predictive uncertainties. However, ground-truth labels are often costly to obtain, which motivates us to accommodate this more realistic scenario. This is relevant because even if the detector trainer may have access to ground-truth labels when learning detectors, these ground-truth labels may not be available to those that aim to quantify the detectors' uncertainties in making predictions.

\subsubsection{Selecting Metrics When Ground-Truth Labels Are Given}

There are standard metrics that can be applied for this purpose. However, these metrics treat each example as equally important and are insensitive to data imbalance,
which is often encountered in malware detection.
This prompts us to propose variants of standard metrics to deal with the issue of data imbalance.
Specifically, we use the following three predictive uncertainty metrics, which are applicable when ground-truth labels are known.

The first standard metric is known as Negative Log-Likelihood (NLL), which are commonly used as a loss function for training a model and intuitively measure the goodness of a model fitting the dataset \cite{DBLP:conf/mlcw/CandelaRSBS05}.
A smaller NLL value means a better calibration.
Formally, this metric is defined as: $$\mathbb{E}_{(z,y)\in D_{test}}-\left(y\log p(y=1|z,\theta)+(1-y)\log(1-p(y=1|z,\theta))\right).$$
In order to deal with imbalanced data, we propose using the following variant, dubbed Balanced NLL (bNLL), which is formally defined as: $\frac{1}{|\mathcal{Y}|}\sum_{i=0}^{|\mathcal{Y}|}\NLL(D_{test}^i)$, 
where $\NLL(D_{test}^i)$ is the expectation of negative log-likelihood on the test set $D_{test}^i$
of class $i$, $i\in \{0,1\}$. bNLL treats each class equally important, while NLL treats each sample equally.

The second standard metric is known as Brier Score Error (BSE), which measures the accuracy of probabilistic predictions \cite{brier1950verification,snoek2019can}. 
A smaller BSE value means a better calibration.
This metric is formally defined as: $\mathbb{E}_{(z, y)\in D_{test}} (y - p(y=1|z,\theta))^2$. 
In order to deal with imbalanced data, we propose using the following variant, dubbed balanced BSE (bBSE), which is formally defined as: 
$\frac{1}{|\mathcal{Y}|}\sum_{i=0}^{|\mathcal{Y}|}\brier(D_{test}^i)$, where $\brier(D_{test}^i)$ is the expectation of BSE on the test set $D_{test}^i$ of class $i$, $i\in\{0,1\}$.

The third standard metric is known as Expected Calibration Error (ECE), which also measures accuracy of predicted probabilities yet in a fine-grained manner \cite{naeini2015obtaining}.
A smaller ECE value means a better calibration. 
Formally, this metric is defined as follows.
Given $S$ buckets corresponding to quantiles of the probabilities $\{\rho_i\}_{i=1}^{S}$, the ECE is defined as
\begin{equation}
	\text{ECE} = \sum_{s=1}^{S}\frac{B_s}{N}|\Pr(y=1|B_s) - \text{conf}(B_s)|, \label{eq:ece}
\end{equation}
where with a little abuse of notation, $B_s=\{(z,y)\in D_{test}:p(y=1|z,\theta)\in(\rho_s,\rho_{s+1}]\}$ is the number of examples in bucket $s$, $\Pr(y=1|B_s)=|B_s|^{-1}\sum_{(z,y)\in B_s}[y = 1]$ is the fraction of malicious examples, $\text{conf}(B_s)=|B_s|^{-1}\sum_{(z,y)\in B_s}{p(y=1|z,\theta)}$, and $N=|D_{test}|$. 
The ECE metric suffers from imbalanced data because the majority class dominates some bins owing to the weights $B_s/N$. 
In order to deal with imbalanced data, we propose using the following variant, dubbed Unweighted ECE (uECE), which is formally defined as follows. $$\text{uECE} = \sum_{s=1}^{S}\frac{1}{S}|\Pr(y=1|B_s) - \text{conf}(B_s)|.$$

\subsubsection{Selecting Metrics When Ground-Truth Labels Are Not Given}

There are three metrics  \cite{kendall2017uncertainties,pearce2020uncertainty,lakshminarayanan2017simple} that can be applied to quantify predictive uncertainty in the absence of ground-truth labels.  These metrics are applied to a point (i.e., expectation is not considered), which indicates these metrics do not suffer from the imbalanced dataset.

The first metric is known as Entropy, which intuitively measure a state of disorder in a physical system \cite{kendall2017uncertainties}. A larger entropy value means a higher uncertainty; 
Formally, this metric is defined as:
\begin{eqnarray}
	-(p(y|z,\theta)\log p(y|z,\theta) +  (1-p(y|z,\theta))\log(1-p(y|z,\theta))), \label{eq:entropy}
\end{eqnarray}
where $p(y|z,\theta)$ denotes the model output $p(y=1|z,\theta)$ or Eq.\eqref{map-pred}. 

The second metric is known as Standard Deviation (SD), which intuitively measures the inconsistency between the base classifiers and the ensemble one \cite{pearce2020uncertainty}. 
A larger SD value means a higher uncertainty.
Formally, this metric is defined as: $$\sqrt{\frac{T}{T-1}\sum_{i=1}^{T}w_i\left(p(y|z,\theta^i) - p(y|z,\theta)\right)^2},$$ where $\theta^i$ denotes the $i$-th DNN model, which has a weight $w_i$ (ref. Eq.\eqref{eq:wens}) or $w_i=1/T$. 

The third metric is known as the KL divergence, which is an alternative to SD \cite{lakshminarayanan2017simple}. A larger KL value means a higher uncertainty.
Formally, this metric is defined as:
$$\sum_{i=1}^T w_i (\kl(p(y|z, \theta^i)||p(y|z,\theta))),$$ where $\kl$ denotes the Kullback-Leibler divergence.

\subsection{Answering RQs}

At this point, one can apply the calibrated detectors to quantify their predictive uncertainties. 
It should be clear that this methodology can be adopted or adapted by other researchers to conduct empirical studies with more kinds of detectors and more metrics. 

\section{Experimental Results and Analysis} \label{sec:experiments}

\subsection{Experimental setup} \label{sec:exp-setup}
We implement the framework using TensorFlow \cite{abadi2016tensorflow} and TensorFlow Probability libraries \cite{dillon2017tensorflow} and run experiments on a CUDA-enabled GTX 2080 Ti GPU. We below detail datasets and hyperparameters.

\subsubsection{Datasets} 

We use 3 widely-used Android datasets: Drebin \cite{arp2014drebin}, VirusShare \cite{VirusShare:Online}, and Androzoo \cite{Allix:2016:ACM:2901739.2903508}. 

\noindent{\bf Drebin}: The Drebin dataset is built before the year of 2013 and is prepossessed by a recent study \cite{li2020adversarial}, which relabels the APKs using the VirusTotal service \cite{VirusTotal:Online} that contains more than 70 antivirus scanners. An APK is treated as malicious if four or more scanners say it is malicious, and benign if no
scanners say it is malicious; theoretical justification of such heuristic but widely-used practice has yet to be made \cite{DBLP:journals/tifs/DuSCCX18}. The resultant Drebin dataset contains 5,560 malicious APKs and 42,333 benign APKs.

\begin{figure}[!b]
	\centering
	\includegraphics[width=0.399\textwidth]{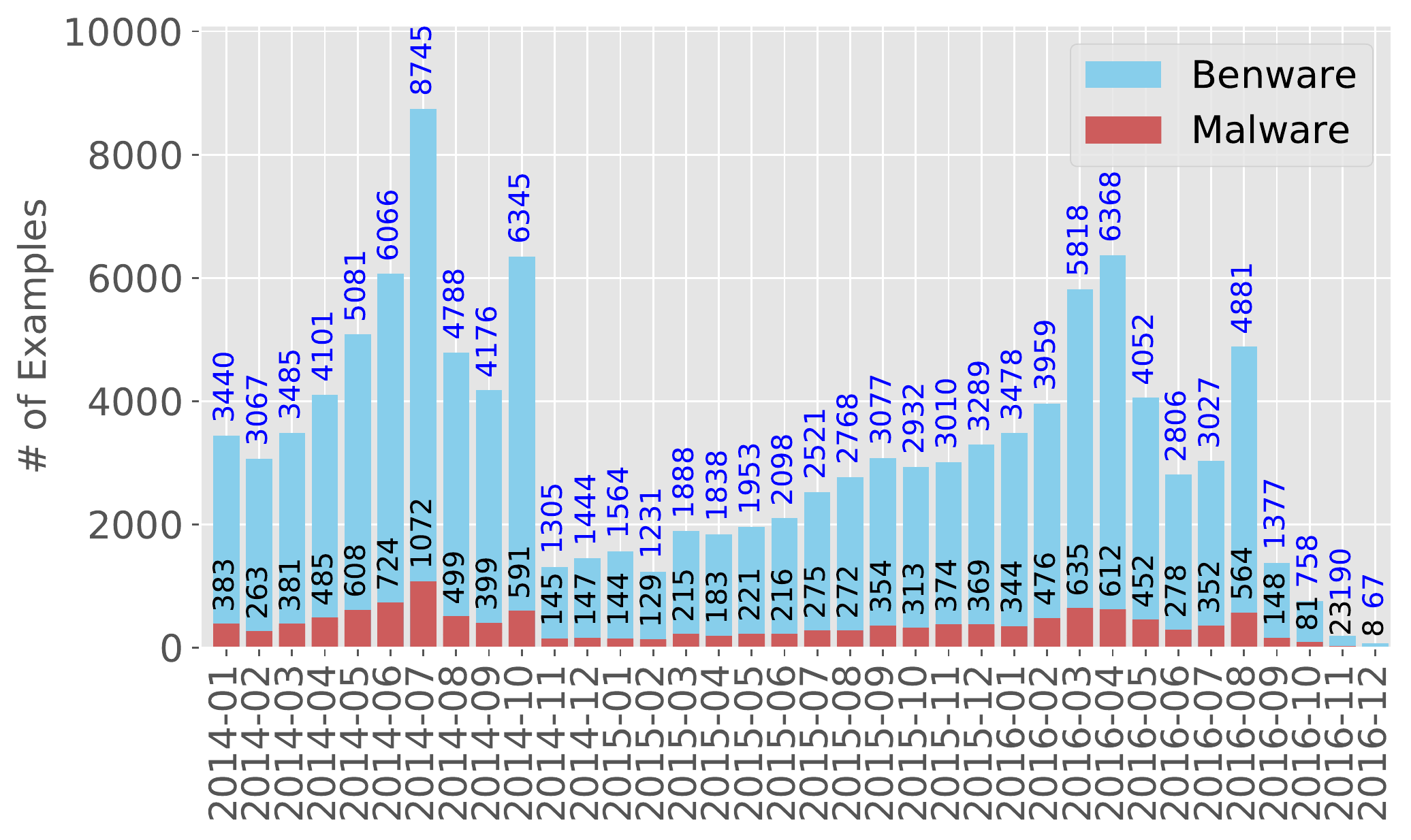}
	\caption{The Androzoo examples of APKs with time \cite{235493}.}
	\label{fig:androzoo-dataset}
\end{figure}

\noindent
{\bf VirusShare}: VirusShare is a repository of potential malware. We chose the APKs collected in 2013 and treat this dataset as {\em out-of-source} in regards to the Drebin dataset. These APKs are labeled using the same fashion as the Drebin dataset, producing 12,383 malicious APKs and 340 benign APKs, while throwing away the 469 APKs in the grey area (i.e., at least one, but at most three, scanners say they are malicious). 

\begin{table*}[!t]
	\caption{Detection estimation and uncertainty evaluation of calibrated malware detectors in the absence of dataset shift.}
	\centering
	\begin{tabular}{c|l|ccccc|cc|cc|cc}
		\toprule
		\multirow{2}{*}{\specialcell{Malware \\ detector}}& \multirow{2}{*}{\specialcell{Calibration \\ method}} & \multicolumn{5}{c|}{Detection estimation (\%)} & \multicolumn{6}{c}{Uncertainty evaluation} \\\cmidrule{3-13}
		& &FNR&FPR&Acc&bAcc&F1&NLL&bNLL&BSE&bBSE&ECE&uECE \\ 
		\midrule\midrule
		
		\multirow{6}{*}{DeepDrebin}&Vanilla&3.90 & 0.31 & 99.28 &	97.90 & 96.84 
		& 0.100 & 0.329 
		& 0.007 & 0.020 
		& 0.006 & 0.104 
		\\
		& Temp scaling & 3.90 & 0.31 & 99.28 & 97.90 & 96.84
		& 0.052 & 0.109
		& 0.006 & 0.018
		& 0.007 & 0.062 
		\\
		& MC Dropout & 3.45 & 0.32 & 99.32 & {\bf 98.12} & 97.04
		& {\bf 0.033} & {\bf 0.094} 
		& 0.006 & {\bf 0.015}
		& {\bf 0.002} & {\bf 0.056} 
		\\
		& VBI & {\bf 3.36} & 0.83 & 98.88 & 97.91 & 95.22
		& 0.054 & {\bf 0.094}
		& 0.009 & 0.016
		& 0.012 & 0.102
		\\
		& Ensemble & 3.99 &	{\bf 0.19} & {\bf 99.37} & 97.91 & {\bf 97.24} 
		& 0.063 & 0.211
		& {\bf 0.005} & 0.018
		& 0.005 & 0.160
		\\
		& wEnsemble & 3.99 & 0.20 &	99.36 &	97.90 & 97.20
		& 0.058 & 0.190
		& {\bf 0.005} & 0.018
		& 0.004 & 0.095 
		\\
		\midrule
		\multirow{6}{*}{MultimodalNN}&Vanilla& {\bf 2.16} & 0.63 & 99.20 & {\bf 98.61} & 96.58
		& 0.087 & 0.207 
		& 0.007 & 0.013
		& {0.007} & 0.162 
		\\
		& Temp scaling & {\bf 2.16} & 0.63 & 99.20 & {\bf 98.61} & 96.58
		& 0.053 & {\bf 0.086} 
		& 0.007 & {\bf 0.012}
		& 0.010 & 0.182 
		\\
		& MC Dropout & 2.97 & 0.39 & 99.31 & 98.32 & 97.03
		& 0.077 & 0.225 
		& {\bf 0.005} & 0.014
		& 0.003 & 0.072 
		\\
		& VBI & 8.45 & 0.33 & 98.73 & 95.61 & 94.35
		& 0.073 & 0.253
		& 0.010 & 0.036
		& 0.004 & 0.078 
		\\
		& Ensemble & 2.52 &	0.39 & 99.36 & 98.55 & 97.26
		& 0.063 & 0.188
		& {\bf 0.005} & {\bf 0.012}
		& 0.004 & 0.144 
		\\
		& wEnsemble & 2.88 & {\bf 0.26} & {\bf 99.44} & 98.43 & {\bf 97.56}
		& {\bf 0.046} & 0.153
		& {\bf 0.005} & 0.013
		& {\bf 0.002} & {\bf 0.045} 
		\\
		\midrule
		\multirow{6}{*}{DeepDroid}&Vanilla& 8.53 & 0.69 &	98.41 & 95.39 &	92.99
		& 0.097 & 0.311
		& 0.013 & 0.039
		& 0.009 & 0.059
		\\
		& Temp scaling & 8.53 & 0.69 &	98.41 &	95.39 & 92.99
		& 0.076 & 0.220
		& 0.013 & 0.038
		& {\bf 0.002} & {\bf 0.023}
		\\
		& MC Dropout & 7.62 & 0.66 & 98.54 & 95.86 & 93.57
		& 0.078 & 0.239
		& 0.012 & 0.035 
		& 0.004 & 0.055
		\\
		& VBI & 12.7 & 0.35 & 98.22 & 93.47 & 91.88
		& 0.084 & 0.305
		& 0.014 & 0.052
		& 0.010 & 0.092
		\\
		& Ensemble & 7.44 &	{\bf 0.11} & 99.05 & 96.23 & 95.73
		& {\bf 0.049} & 0.171
		& 0.008 & 0.029
		& 0.006 & 0.162
		\\
		& wEnsemble & {\bf 4.99} & 0.27 & {\bf 99.18} & {\bf 97.37} & {\bf 96.41}
		& 0.050 & {\bf 0.131}
		& {\bf 0.007} & {\bf 0.022}
		& 0.008 & 0.066
		\\
		\midrule
		\multirow{6}{*}{Droidetec}&Vanilla& 4.14 & 1.53 &	98.17 &	97.17 & 92.30
		& 0.119 & 0.208
		& 0.016 & 0.026
		& 0.014 & 0.205
		\\
		& Temp scaling & 4.14 & 1.53 &	98.17 &	97.17 & 92.30
		& 0.101 & 0.167
		& 0.016 & 0.026
		& 0.015 & 0.180
		\\
		& MC Dropout & {\bf 3.03} & 1.40 &	98.41 &	97.78 & 93.32
		& 0.088 & 0.142
		& 0.014 & 0.020
		& 0.013 & 0.207
		\\
		& VBI & 3.22 & 1.67 & 98.15 & 97.55 & 92.29
		& 0.118 & 0.202
		& 0.015 & 0.022
		& 0.015 & 0.262
		\\
		& Ensemble & 3.13 & 0.75 & 98.98 & {\bf 98.06} & 95.60
		& 0.055 & 0.107
		& 0.009 & {\bf 0.016}
		& 0.010 & 0.143
		\\
		& wEnsemble & 3.49 & {\bf 0.59} & {\bf 99.07} & 97.96 & {\bf 95.98}
		& {\bf 0.046} & {\bf 0.101}
		& {\bf 0.007} & {\bf 0.016}
		& {\bf 0.008} & {\bf 0.116}
		\\
		\bottomrule
	\end{tabular}
	\label{tab:drebin-effectiveness}
\end{table*}

\noindent
{\bf Androzoo}: Androzoo is an APK repository, including over 10 million examples, along with their VirusTotal reports. These APKs were crawled from the known markets (e.g., Google Play, AppChina). Following a previous study \cite{235493}, we use a subset of these APKs spanning from January 2014 to December 2016, which includes 12,735 malicious examples and 116,993 benign examples. Figure \ref{fig:androzoo-dataset} plots the monthly distribution of these examples with time \cite{235493}. 

\subsubsection{Hyper-parameters} We present the hyper-parameters for malware detectors and then for calibration methods. 

\noindent{\bf DeepDrebin} \cite{grosse2017adversarial} is an MLP, consisting of two fully-connected hidden layers, each with 200 neurons. 

\noindent{\bf MultimodalNN} \cite{8443370} has five headers and an integrated part, where each header consists of two fully-connected layers of 500 neurons. The integrated part consists of two fully-connected layers of 200 neurons. 

\noindent{\bf DeepDroid} \cite{10.1145/3029806.3029823} has an embedding layer, followed by a convolutional layer and two fully-connected layers. The vocabulary of the embedding layer has 256 words, which correspond to 256 {\em Dalvik} opcodes; the embedding dimension is 8; the convolutional layer has the kernels of size 8$\times$8 with 64 kernels; the fully-connected layer has 200 neurons. Limited by the GPU memory, DeepDroid can only deal with a maximum sequence of length 700,000, meaning that APKs with longer opcode sequences are truncated and APKs with shorter opcode sequences are padded with 0's (which correspond to {\tt nop}). 

\noindent{\bf Droidetec} \cite{ma2020droidetec} has an embedding layer with vocabulary size 100,000 and embedding dimension 8, the bi-directional LSTM layer with 64 units, and a fully-connected hidden layer with 200 neurons. Droidetec allows the maximum length of API sequence 100,000. We further clip the gradient values into the range of [-100, 100] for Droidetec in case of gradient explosion.

All models mentioned above use the ReLU activation function. We also place a dropout layer with a dropout rate 0.4 before the last fully-connected layer (i.e., the output layer).
We implement the four malware detectors by ourselves. 
The hyperparameters of calibration methods are detailed below.

\noindent
{\bf Vanilla} and {\bf Temp scaling}: We have the same settings as the malware detectors mentioned above. 

\noindent
{\bf MC dropout}: We use a dropout layer with a dropout rate 0.4. We add the dropout layer into the fully-connected layer, convolutional layer, and the LSTM layer, respectively. Following a recent study \cite{snoek2019can}, we neglect the $\ell_2$ regularization that could decline the detection accuracy. In the test phase, we sample 10 predictions for each example.

\noindent
{\bf VBI}: We sample the parameters of the fully-connected layer or the convolutional layer (i.e., weights and bias) from Gaussian distributions. A Gaussian distribution has the variables ({\em mean} and {\em standard deviation}), which are learned via back propagation using the reparameterization technique \cite{blundell2015weight}. We do not implement VBI for Bi-LSTM, due to the effectiveness issue \cite{snoek2019can,DBLP:conf/iclr/WangLN21}. This means only the last layer of Droidetec is calibrated by VBI. In the test phase, we sample 10 predictions for each example.

\noindent
{\bf Ensemble} and {\bf wEnsemble}: We learn 10 base instances for each ensemble-based method.

We learn these models using the Adam optimizer with 30 epochs, batch size 16, and learning rate 0.001. A model is selected for evaluation when it achieves the highest accuracy on the validation set in the training phase. In addition, we calculate the validation accuracy at the end of each epoch.

\subsection{Answering RQ1}

In order to quantify the predictive uncertainty of malware detectors in the absence of dataset shift, we learn the aforementioned 24 malware detectors on the Drebin dataset, by splitting it into three disjoint sets of 60\% for training, 20\% for validation, and 20\% for testing. 

Table \ref{tab:drebin-effectiveness} summarizes the results, including detection estimation using the metrics False Negative Rate
(FNR), False Positive Rate (FPR), Accuracy (Acc) or percentage of detecting benign and malicious samples correctly, balanced Accuracy (bAcc) \cite{5597285} and F1 score \cite{Pendleton:2016}, and uncertainty evaluation using the metrics NLL, bNLL, BSE, bBSE, ECE, and uECE. We make four observations. 
First, {\em Temp scaling} achieves the same detection accuracy as its vanilla model because it is a post-processing method without changing the learned parameters. On the other hand, MC dropout, Ensemble and wEnsemble improve detection accuracy but VBI degrades detection accuracy somewhat when compared with the vanilla model. In terms of balanced accuracy, the calibration methods do not always improve detection accuracy because the vanilla MultimodalNN actually achieves the highest balanced accuracy 98.61\% among all those detectors. The reason may be that MultimodalNN itself is an ensemble model (e.g., 5 headers are equipped).

Second, ensemble methods (i.e., Ensemble and wEnsemble) reduce the calibration error when compared with the respective vanilla models. Moreover, the two ensemble methods exceed the other calibration methods in terms of the NLL, BSE and bBSE metrics, except for DeepDrebin (suggesting MC Dropout is best for calibration). Third, the measurements of the balanced metrics are notably larger than their imbalanced counterparts (e.g., bNLL vs. NLL), because benign examples dominate the test set and malware detectors predict benign examples more accurately than predicting malicious ones. 

\begin{figure}[!htbp]
    \vspace{-8pt}
	\centering
	\subfloat[\# of examples per bin.]{
	    \includegraphics[width=0.239\textwidth]{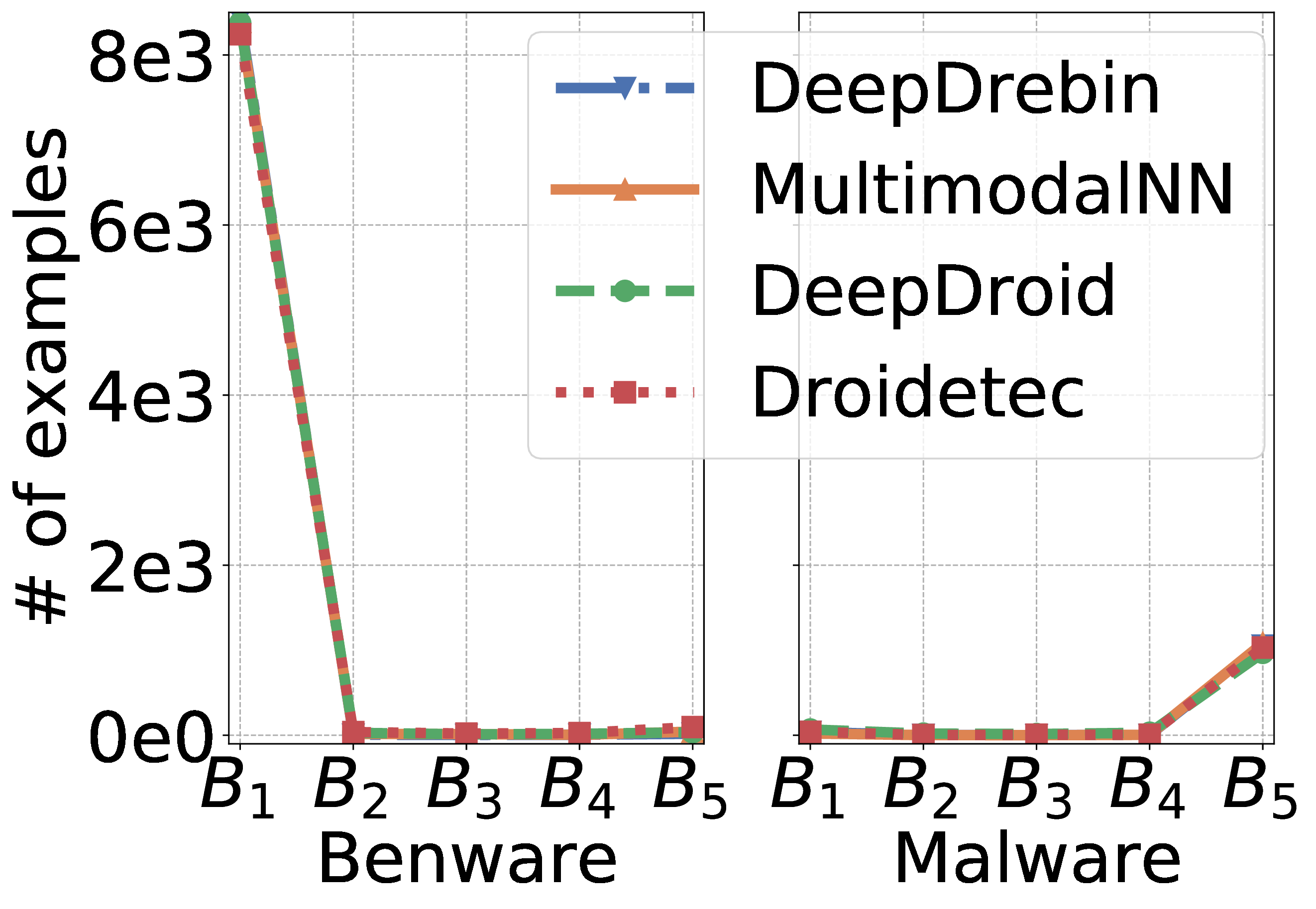} %
	    \label{subfig:rd-02}%
	}
	\subfloat[Reliability diagram]{
	    \includegraphics[width=0.223\textwidth]{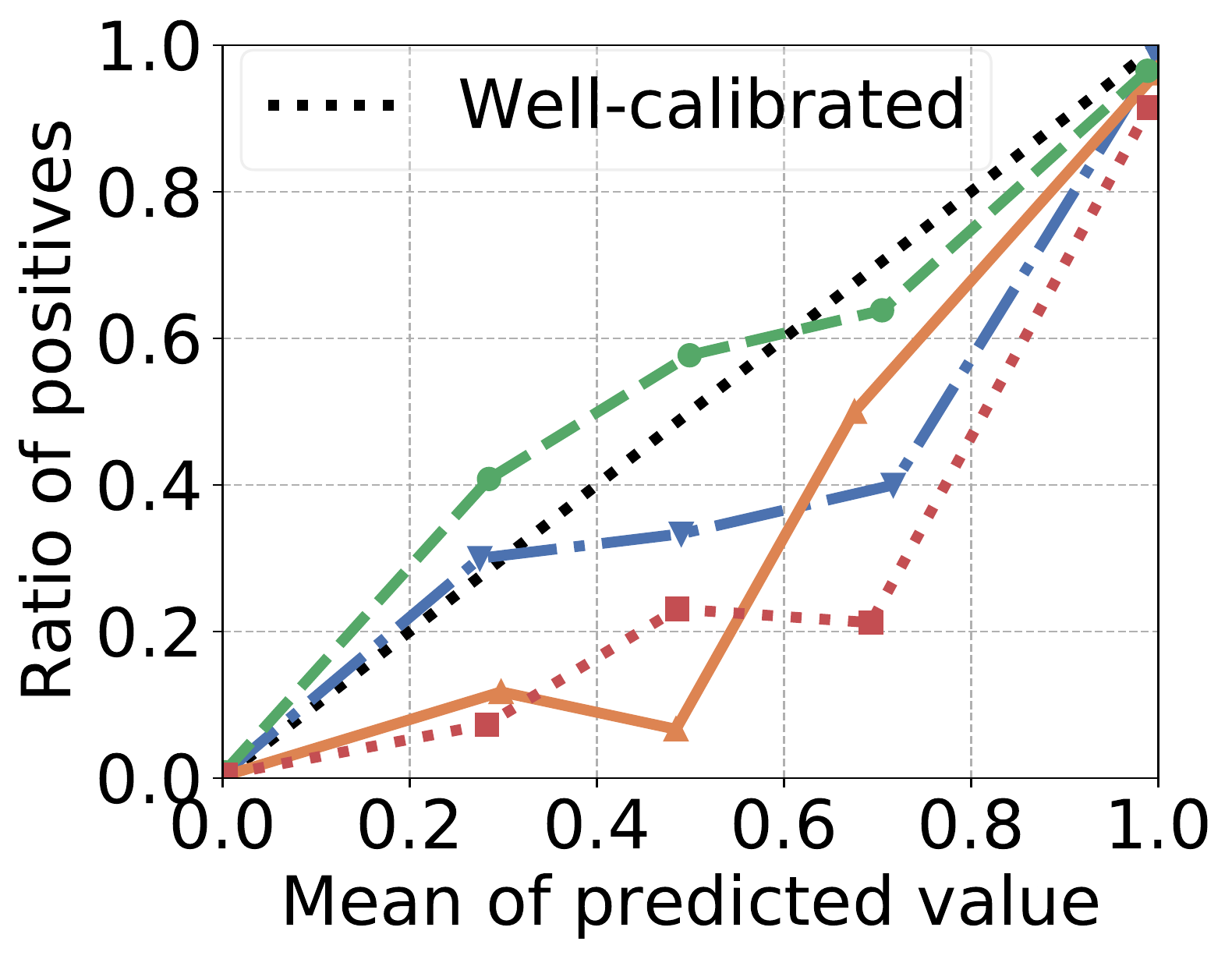}
	    \label{subfig:rd-01}
	}
	\caption{
		Reliability diagram of vanilla malware detectors. There are 5 bins $B_1,\ldots,B_5$ and ``Benware'' denotes the {{\em ben}}ign soft{{\em ware}}.
	}
	\label{fig:reliable-diam}
\end{figure}

Fourth, uECE shows inconsistent results in terms of bNLL and bBSE. In order to understand the reasons, we plot the {\em reliability diagram} \cite{niculescu2005predicting}, which demonstrates the difference between the fraction of malicious examples in each bin, namely the difference between the $\Pr(y=1|B_s)$ in Eq.\eqref{eq:ece} and the mean of the predicted confidence $conf(B_s)$. Figure \ref{fig:reliable-diam} plots the results of the vanilla malware detectors, along with the number of examples in the bins.
Figure \ref{subfig:rd-02} shows that most examples belong to bins $B_1$ and $B_5$. Figure \ref{subfig:rd-01} says DeepDroid achieves the lowest error (because it is closest to the diagonal than others), and shall be best calibrated, which contracts the ECE values in Table \ref{tab:drebin-effectiveness} (demonstrating that MultimodalNN is best instead). This is because as shown in Figure \ref{subfig:rd-02}, most benign examples belong to bin $B_1$ and most malicious examples belong to bin $B_5$. As a comparison, uECE does not suffer from this issue.

\begin{insight}
	Calibration methods reduce malware detection uncertainty; variational Bayesian inference degrades detection accuracy and F1 score in the absence of dataset shift; balanced metrics (i.e., bNLL, bBSE and uECE) are more sensitive than their imbalanced counterparts (i.e., NLL, BSE and ECE) when data imbalance is present.
\end{insight}

\begin{figure*}[!htbp]
    \vspace{-8pt}
	\centering
	\subfloat[]{%
        \includegraphics[width=0.7\textwidth]{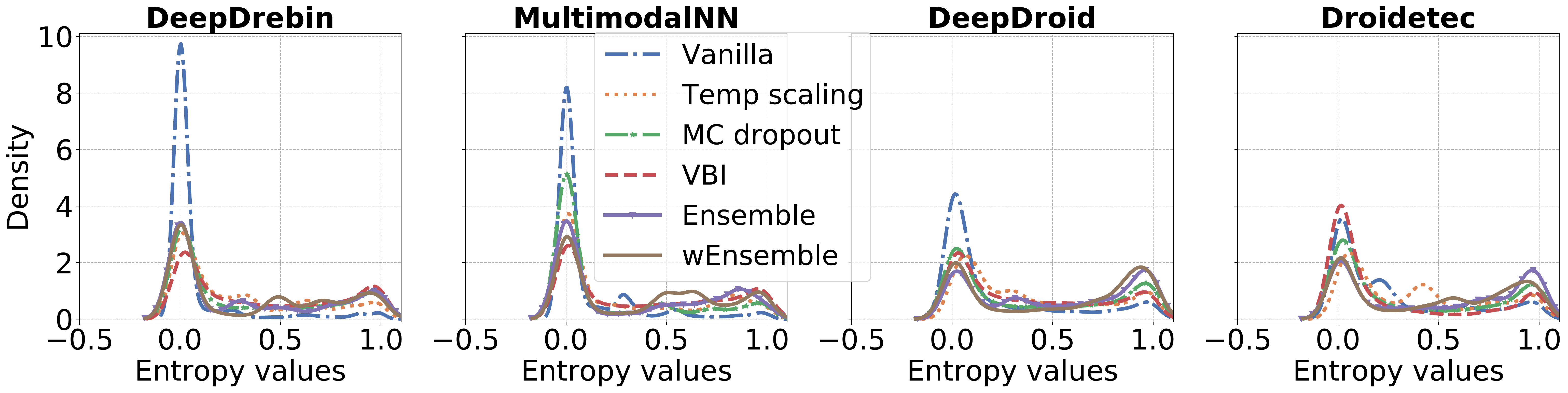}%
        \label{subfig:oos-01} %
    }
    
    \vspace{-8pt}
    \subfloat[]{%
        \includegraphics[width=0.72\textwidth]{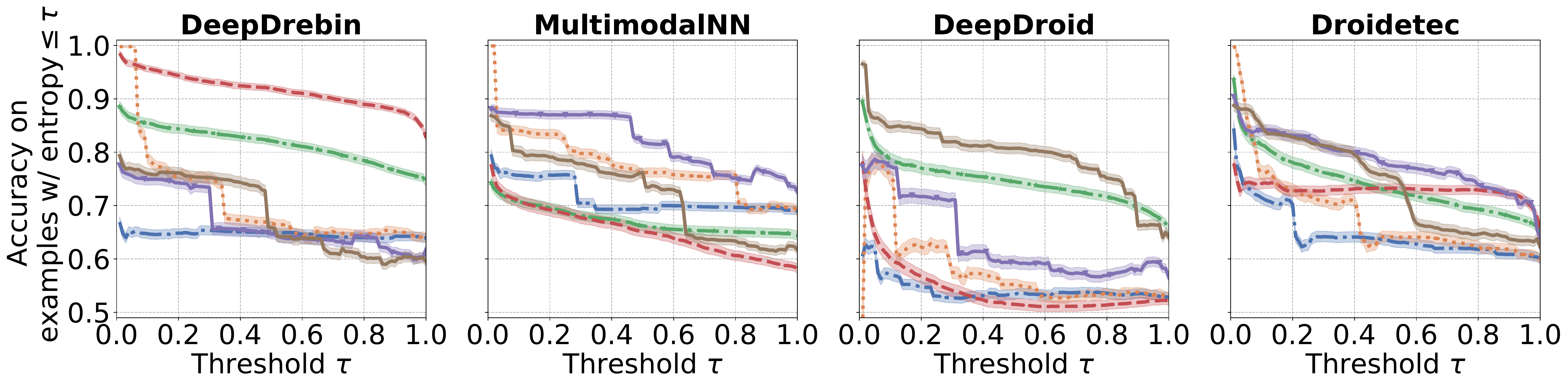}%
        \label{subfig:oos-02} %
    }
    \vspace{-8pt}
	\caption{Predictive entropy (see Eq.\eqref{eq:entropy}) of malware detectors trained on the Drebin dataset and tested on the VirusShare: (a) histogram of predictive entropy; (b) accuracy on the dataset after excluding the examples for which the detector has high uncertainties (i.e., the examples for which the predictive entropy is above a pre-determined threshold $\tau$), which corresponds to the real-world use of the quantified predictive uncertainty.}
	\label{fig:oos-accuracy}
\end{figure*}


\begin{table}[!t]
	\caption{Predictive uncertainty of malware detectors that are learned on the Drebin dataset and tested on the VirusShare.}
	\centering
	\resizebox{0.9\textwidth}{!}{\begin{minipage}{\textwidth}
		\setlength{\tabcolsep}{2.7pt}
		\begin{tabular}{c|l|cc|cc|cc|cc}
			\toprule
			\multirow{2}{*}{} & \multirow{2}{*}{\specialcell{Calibration \\ method}} & \multicolumn{2}{c|}{Detection(\%)} & \multicolumn{6}{c}{Uncertainty evaluation} \\\cmidrule{3-10}
			& &Acc&bAcc&NLL&bNLL&BSE&bBSE&ECE&uECE \\ 
			\midrule\midrule
			
			\multirow{6}{*}{\vthead{DeepDrebin}}&Vanilla & 63.96 & 77.59
			& 5.52 & 3.58
			& 0.35 & 0.21
			& 0.359 & 0.483
			\\
			& Temp scaling & 63.96 & 77.59
			& 1.65 & 1.39
			& 0.31 & 0.19
			& 0.363 & 0.468
			\\
			& MC Dropout & 74.39 & 83.53
			& 1.48 & 0.95
			& 0.20 & 0.13
			& 0.274 & 0.463
			\\
			& VBI & {\bf 82.69} & {\bf 87.07}
			& {\bf 0.64} & {\bf 0.52}
			& {\bf 0.13} & {\bf 0.10}
			& {\bf 0.210} & {\bf 0.434}
			\\
			& Ensemble & 61.76 & 78.04
			& 3.43 & 2.15
			& 0.32 & 0.19
			& 0.394 & 0.466
			\\
			& wEnsemble & 59.49 & 76.59
			& 3.08 & 1.79
			& 0.32 & 0.19
			& 0.397 & 0.469
			\\
			\midrule
			\multirow{6}{*}{\vthead{MultimodalNN}}&Vanilla& 69.02 &	80.48
			& 4.05 & 2.59
			& 0.29 & 0.18
			& 0.306 & 0.463
			\\
			& Temp scaling & 69.02 & 80.48
			& {\bf 1.45} & {\bf 1.13}
			& 0.25 & 0.17
			& 0.303 & 0.464
			\\
			& MC Dropout & 64.38 & 78.38
			& 4.61 & 2.58
			& 0.32 & 0.18
			& 0.361 & 0.473
			\\
			& VBI & 58.17 &	77.21
			& 2.32 & 1.25
			& 0.33 & 0.18
			& 0.414 & 0.478
			\\
			& Ensemble & {\bf 72.66} & {\bf 82.20}
			& 2.24 & 1.34
			& {\bf 0.21} & {\bf 0.14}
			& {\bf 0.279} & {\bf 0.450}
			\\
			& wEnsemble & 61.73 & 77.74
			& 2.21 & 1.26
			& 0.30 & 0.17
			& 0.379 & 0.471
			\\
			\midrule
			\multirow{6}{*}{\vthead{DeepDroid}}&Vanilla & 52.85 & 70.02
			& 3.44 & 2.11
			& 0.42 & 0.27
			& 0.463 & 0.481
			\\
			& Temp scaling & 52.85 & 70.02
			& 2.09 & 1.31
			& 0.39 & 0.25
			& 0.460 & 0.477
			\\
			& MC Dropout & {\bf 65.50} & {\bf 74.22}
			& 1.54 & 1.10
			& 0.26 & 0.20
			& {\bf 0.338} & 0.465
			\\
			& VBI & 56.57 & 72.65
			& 2.44 & 1.46
			& 0.40 & 0.25
			& 0.476 & 0.475
			\\
			& Ensemble & 56.57 & 72.65
			& 1.78 & 1.12
			& 0.33 & 0.20
			& 0.435 & 0.475
			\\
			& wEnsemble & 64.11 & {\bf 74.22}
			& {\bf 1.22} & {\bf 1.00} 
			& {\bf 0.24} & {\bf 0.19}
			& 0.350 & {\bf 0.461} 
			\\
			\midrule
			\multirow{6}{*}{\vthead{Droidetec}}&Vanilla &	59.98 & 70.04
			& 2.22 & 1.59
			& 0.34 & 0.253
			& 0.389 & 0.476 
			\\
			& Temp scaling & 59.98 & 70.04
			& 1.66 & 1.20 
			& 0.32 & 0.235
			& 0.390 & 0.476
			
			\\
			& MC Dropout & 64.99 & 72.61
			& 1.57 & 1.33 
			& 0.27 & 0.222 
			& 0.344 & 0.472 
			
			\\
			& VBI & {\bf 65.30} & {\bf 74.21}
			& 2.14 & 1.85
			& 0.27 & 0.210
			& {\bf 0.315} & 0.473 
			
			\\
			& Ensemble & 63.83 & 72.45
			& {\bf 1.33} & {\bf 0.94} 
			& {\bf 0.24} & {\bf 0.185}
			& 0.343 & {}0.470
			
			\\
			& wEnsemble & 62.26 & 71.64
			& 1.50 & 1.23 
			& 0.28 & 0.219
			& 0.375 & 0.465 
			\\	
			\bottomrule
		\end{tabular}
	\end{minipage}}
	\vspace{-10pt}
	\label{tab:drebin-oos-effectiveness}
\end{table}

\subsection{Answering RQ2} \label{sec:aq2}
In order to quantify the predictive uncertainty of malware detectors with respect to out-of-source examples, we apply the Drebin malware detectors to the VirusShare dataset. We assess predictive distribution and report the accuracy of malware detectors on the datasets obtained after removing the examples for which the detectors are uncertain (i.e., with an entropy value above a threshold $\tau$); this corresponds to the real-world usefulness of quantifying predictive uncertainty (i.e., discarding prediction results for which detector is uncertain). 

Table \ref{tab:drebin-oos-effectiveness} summarizes the uncertainty evaluation and the corresponding detection accuracy. We make three observations. 
(i) Malware detectors achieve low accuracy with out-of-source test examples. Nevertheless, DeepDrebin incorporating VBI obtain an accuracy of 82.69\%, which notably outperforms other detectors. It is reminding that VBI hinders the detection accuracy in the absence of dataset shift.
(ii) Calibration methods (e.g., VBI or Ensemble) reduce the uncertainty in terms of bNLL and bBSE when compared with the vanilla models, except for the MultimodalNN model incorporating MC dropout.
(iii) DeepDrebin incorporating VBI also achieves the best calibration results, suggesting that VBI benefits from both regularization and calibration. On the other hand, DeepDroid and Droidetec suffer from the setting of {\em out of source}. Both models handle the very long sequential data that would be truncated due to the limited GPU memory, leading to the inferior results.

\begin{figure}[!t]
	\centering
	\includegraphics[width=0.399\textwidth]{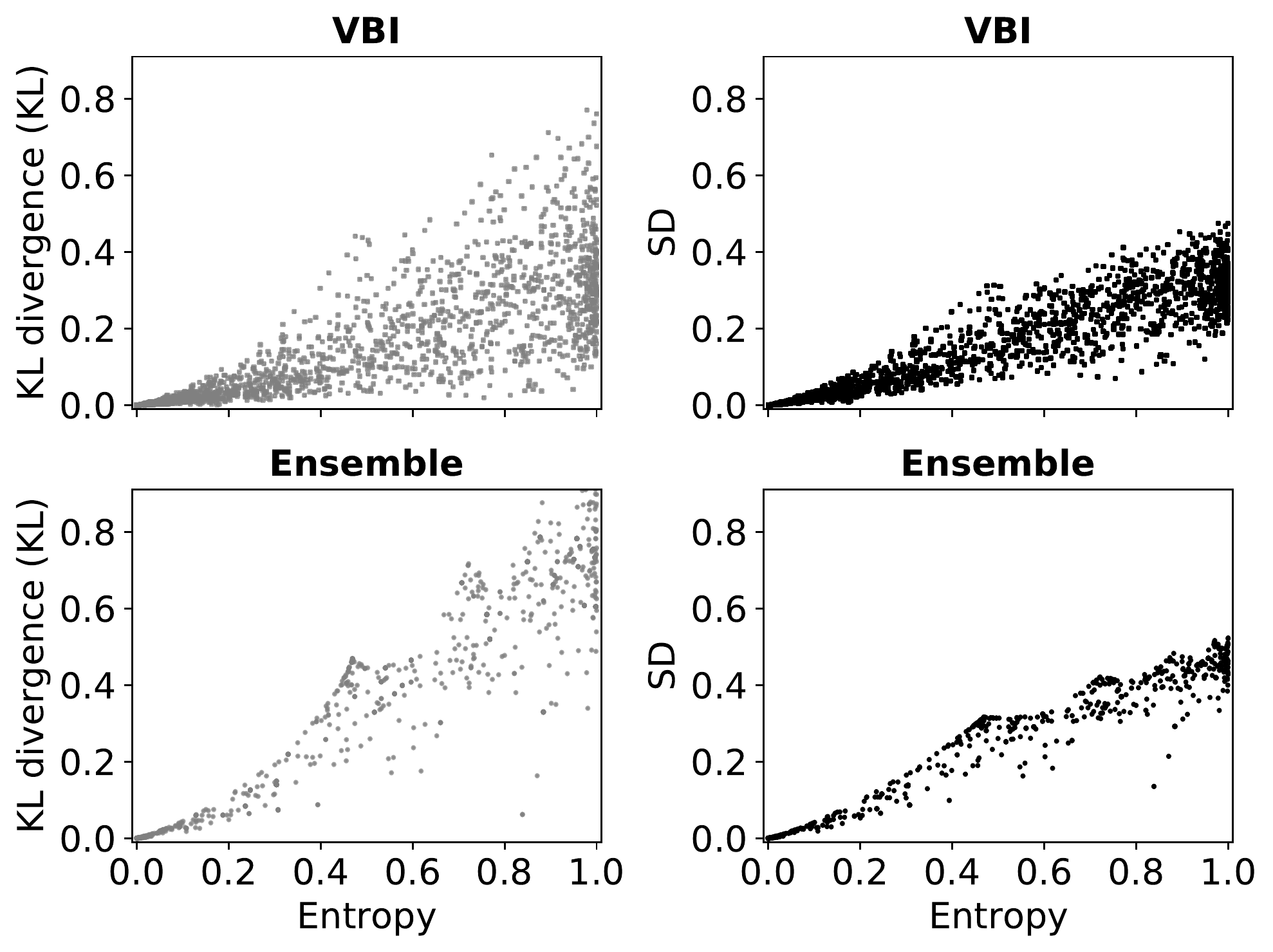}
	\caption{Scatter points of the randomly selected 2,000 test examples from the VirusShare dataset with paired values (Entropy, KL) and (Entropy, SD) that are obtained by applying DeepDrebin incorporating VBI or Ensemble.
	}
	\label{fig:drebin-entropy-kl-std}
\end{figure}

\begin{figure*}[!htbp]
	\centering
	\includegraphics[width=0.85\textwidth]{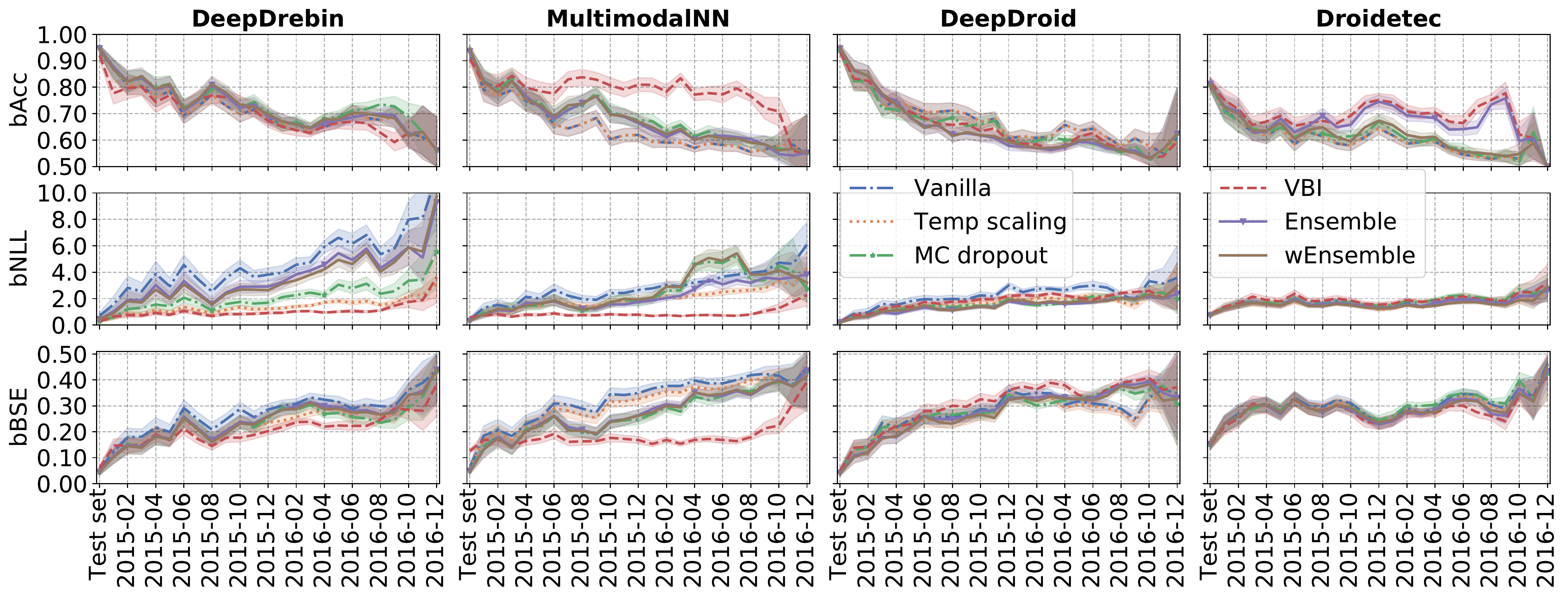}
	\caption{Illustration of balanced accuracy (bAcc), balanced NLL (bNLL), balanced BSE (bBSE) under temporal covariate shift.}
	\label{fig:androzoo-bacc-bnll-bbrier}
\end{figure*}

\begin{figure*}[!tbp]
    \vspace{-10pt}
	\centering
	\subfloat[]{%
      \includegraphics[width=0.67\textwidth]{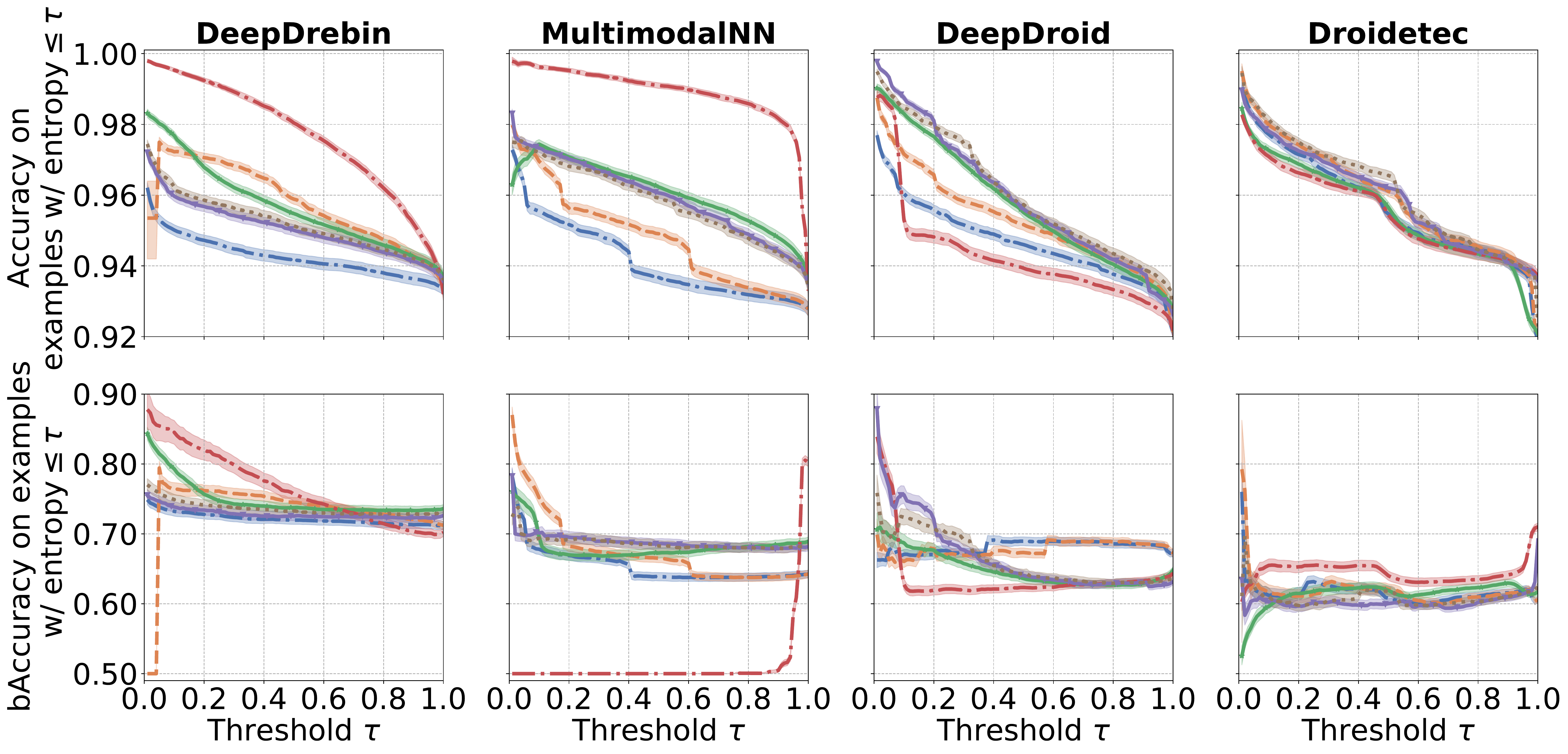}%
      \label{subfig:androzoo-ent-acc-bacc}%
    }
	\subfloat[]{%
      \includegraphics[width=0.315\textwidth]{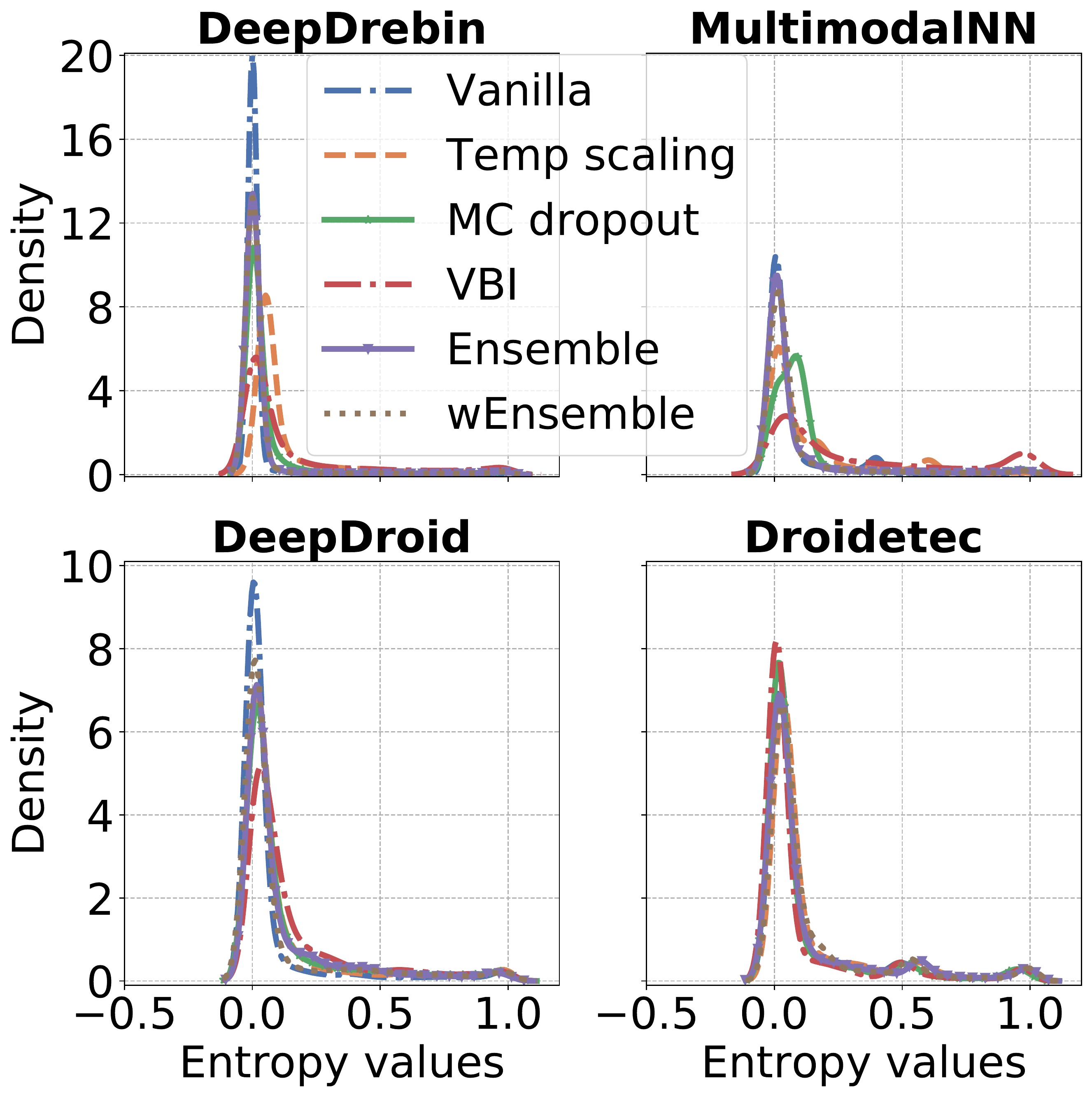}%
      \label{subfig:androzoo-density}%
    }
	\caption{Predictive entropy (see Eq.\eqref{eq:entropy}) of malware detectors on Androzoo dataset: (a) accuracy (upper row) and bAccuracy (bottom row) on the Androzoo test set after excluding the examples for which the detectors have high uncertainties (i.e., the examples for which the predictive entropy is above a pre-determined threshold $\tau$); (b) test sample density of predictive entropy.}
	\label{fig:androzoo-ent}
\end{figure*}



Figure \ref{fig:oos-accuracy} illustrates the density of predictive entropy. Figure \ref{subfig:oos-02} further shows the accuracy on the examples after removing the ones for which the detectors are uncertain about their predictions. We make the following observations.
(i) The vanilla models return zero entropy value for many examples, but calibration methods relieve this situation notably. Considering that the higher entropy delivers more uncertainty, vanilla model is poorly calibrated regarding the out-of-source examples. This is further confirmed that vanilla models exhibit a limited change along with increasing thresholds in Figure \ref{subfig:oos-02}. 
(ii) Ensemble and wEnsemble increase the robustness of deep learning models in detecting out-of-source examples. For example, the MultimodalNN, DeepDroid, and Droidetec models can be enhanced by Ensemble and wEnsemble to some extent, especially when applied to predicting the examples with entropy values below 0.3 (i.e., the detectors are relatively certain about their predictions). 
(iii) DeepDrebin incorporating either VBI or {\em MC dropout} make a significant achievement by comparing with ensemble-based calibrations. Because VBI requires suitable prior distributions, it is challenging to generalize this calibration to all models effectively \cite{blundell2015weight}. (iv) {\em Temp scaling} demonstrates an un-intuitive phenomenon that it achieves almost 100\% accuracy at the start of the curves when applied to DeepDrebin, MultimodalNN and Droidetec, but degrades the accuracy of DeepDroid. The is because {\em Temp scaling} tends to over-approximate the predicted probabilities, resulting in high confidence score for examples, some of which are mis-classified, however. In addition, the balanced accuracy demonstrates the similar experimental results (cf. the appendix materials for details).

Figure \ref{fig:drebin-entropy-kl-std} plots the relationship between the entropy and the KL divergence (KL), and the relationship between the entropy and the Standard Deviation (SD). A scatter point represents a test example based on the paired value. We observe that these three measurements are closely related, which explains why we use the entropy to characterize the predictive uncertainty solely. We only report the results for the calibrated DeepDrebin with VBI and Ensemble, because similar phenomena are observed for the other models. Note that Vanilla and Temp scaling do not apply to KL divergence and standard deviation.

\begin{insight}
	Deep ensemble benefits calibration of malware detectors against out-of-source test examples, but a carefully tuned VBI model could achieve a higher quality of uncertainty than ensemble-based methods; Measurements of entropy, KL divergence and standard deviation are closely related.
\end{insight} 

\begin{table*}[!htbp]
	\caption{Effectiveness of calibrated malware detectors under adversarial evasion attacks.
	}
	\centering
	\setlength{\tabcolsep}{2.7pt}
	\begin{tabular}{c|l|cccc|cccc|cccc}
		\toprule
		\multirow{2}{*}{\specialcell{Malware \\ detector}}& \multirow{2}{*}{\specialcell{Calibration \\ method}} & \multicolumn{4}{c|}{No attack} & \multicolumn{4}{c|}{ ``Max'' PGDs+GDKDE attack} & \multicolumn{4}{c}{ Mimicry attack }\\\cmidrule{3-14}
		& & Acc (\%) & NLL & BSE & ECE & Acc (\%) & NLL & BSE & ECE & Acc (\%) & NLL & BSE & ECE \\ 
		\midrule\midrule
		
		\multirow{6}{*}{DeepDrebin} & Vanilla 
		& 96.09 & 0.629 & 0.037 & 0.039
		& {\bf 0.00} & 33.22 & 1.000 & 1.000
		& 66.09 & 4.778 & 0.317 & 0.334
		\\
		& Temp scaling
		& 96.09 & 0.184 & 0.033 & 0.042 
		& {\bf 0.00} & {\bf 7.015} & {\bf 0.985} & {\bf 0.992}
		& 66.09 & 1.427 & 0.266 & 0.332 	
		\\
		& MC Dropout
		& {\bf 96.55} & 0.186 & 0.029 & {\bf 0.040}
		& {\bf 0.00} & 33.22 & 1.000 & 1.000
		& 69.18 & 1.639 & 0.245 & {\bf 0.317}
		\\
		& VBI
		& 96.27 & {\bf 0.142} & {\bf 0.025} & 0.051 
		& {\bf 0.00} & 33.22 & 1.000  & 1.000
		& {\bf 69.91} & {\bf 1.034} & {\bf 0.211} & 0.320 
		\\
		& Ensemble
		& 96.00 & 0.403 & 0.034 & 0.042
		& {\bf 0.00} & 33.22 & 1.000 & 1.000
		& 64.82 & 3.296 & 0.295 & 0.363 
		\\
		& wEnsemble
		& 96.00 & 0.362 & 0.034 & 0.042
		& {\bf 0.00} & 33.22  & 1.000 & 1.000
		& 64.64 & 2.944 & 0.296 & 0.362 	
		\\
		\midrule
		\multirow{6}{*}{\specialcell{Multimo\\-dalNN}}&Vanilla
		& {\bf 97.82} & 0.368 & 0.020 & {\bf 0.023} 
		& {\bf 0.00} & 33.22 & 1.000 & 1.000
		& 87.64 & 1.530 & 0.107 & {\bf 0.119}
		\\
		& Temp scaling
		& {\bf 97.82} & {\bf 0.129} & {\bf 0.019} & 0.025
		& {\bf 0.00} & {\bf 8.871} & {\bf 0.996} & {\bf 0.998}
		& 87.64 & 0.562 & 0.094 & 0.122
		\\
		& MC Dropout
		& 97.18 & 0.399 & 0.024 & 0.030
		& {\bf 0.00} & 33.22 & 1.000 & 1.000
		& 85.64 & 1.822 & 0.121 & 0.152 
		\\
		& VBI
		& 96.82 & 0.166 & 0.026 & 0.042
		& {\bf 0.00} & 33.22 & 1.000 & 1.000
		& {\bf 89.64} & {\bf 0.506} & {\bf 0.080} & {\bf 0.119} 
		\\
		& Ensemble
		& 97.45 & 0.355 & 0.021 & 0.026
		& {\bf 0.00} & 33.22 & 1.000 & 1.000
		& 88.09 & 1.148 & 0.091 & 0.129
		\\
		& wEnsemble
		& 97.09 & 0.295 & 0.025 & 0.035
		& {\bf 0.00} & 33.22 & 1.000 & 1.000
		& 84.27 & 1.124 & 0.123 & 0.187
		\\
		\midrule
		\multirow{6}{*}{DeepDroid}&Vanilla
		& 91.55 & 0.587 & 0.073 & 0.095
		& 85.45 & 0.773 & 0.116 & 0.148
		& 86.09 & 0.786 & 0.110 & 0.142
		\\
		& Temp scaling
		& 91.55 & 0.404 & 0.069 & 0.113
		& 85.45 & 0.536 & 0.105 & 0.165
		& 86.09 & 0.538 & 0.101 & 0.158
		\\
		& MC Dropout
		& 92.55 & 0.451 & 0.066 & 0.097
		& {\bf 93.55} & {\bf 0.273} & {\bf 0.048} & {\bf 0.074}
		& 90.18 & 0.529 & 0.083 & 0.126 
		\\
		& VBI
		& 87.27 & 0.592 & 0.102 & 0.151
		& 84.00 & 0.592 & 0.117 & 0.180
		& 82.00 & 0.705 & 0.136 & 0.200 
		\\
		& Ensemble
		& 92.55 & 0.329 & 0.058 & 0.099
		& 90.64 & 0.366 & 0.068 & 0.129
		& 89.55 & 0.433 & 0.079 & 0.142 
		\\
		& wEnsemble
		& {\bf 95.00} & {\bf 0.237} & {\bf 0.040} & {\bf 0.069}  
		& 93.00 & 0.309 & 0.057 & 0.111
		& {\bf 92.82} & {\bf 0.348} & {\bf 0.061} & {\bf 0.111} 
		\\
		\midrule
		\multirow{6}{*}{Droidetec}&Vanilla
		& 98.08 & 0.123 & 0.016 & 0.028
		& 66.03 & 2.331 & 0.298 & 0.341
		& 88.80 & 0.656 & 0.099 & 0.126 
		\\
		& Temp scaling
		& 98.08 & 0.113 & 0.017 & 0.039
		& 66.03 & 1.717 & 0.285 & 0.345
		& 88.80 & 0.514 & 0.095 & 0.138
		\\
		& MC Dropout
		& 98.81 & 0.063 & 0.009 & 0.018
		& 67.30 & 2.066 & 0.272 & 0.330
		& 93.17 & 0.268 & 0.052 & 0.090
		\\
		& VBI 
		& 98.54 & 0.095 & 0.012 & 0.016 
		& 71.31 & 1.996 & 0.250 & 0.292
		& 93.08 & 0.396 & 0.060 & 0.078
		\\
		& Ensemble
		& {\bf 99.18} & 0.060 & {\bf 0.008} & {\bf 0.014}
		& {\bf 80.87} & {\bf 0.657} & {\bf 0.138} & {\bf 0.215}
		& {\bf 96.17} & 0.175 & 0.030 & 0.068 
		\\
		& wEnsemble
		& 99.00 & {\bf 0.048} & {\bf 0.008} & 0.015
		& 80.05 & 0.776 & 0.151 & 0.232
		& 95.90 & {\bf 0.156} & {\bf 0.029} & {\bf 0.066}
		\\
		\bottomrule
	\end{tabular}
	\label{tab:drebin-adv-attack-full}
\end{table*}

\subsection{Answering RQ3}

In order to quantify the predictive uncertainty of malware detectors under temporal covariate  shift, we use Androzoo dataset. Specifically, we train malware detectors upon the APKs collected in the year of 2014, and test these malware detectors upon APKs collected in the year of 2015 and 2016 at the month granularity. We also split the APKs in the year of 2014 into three disjoint datasets, 83.4\% for training, 8.33\% for validation (8.33\% is the average percentage of APKs emerging in each month of 2014), and 8.33\% for testing. 

Figure \ref{fig:androzoo-bacc-bnll-bbrier} plots the balanced accuracy (bAccuracy), balanced NLL (bNLL) and balanced BSE (bBSE) under temporal covariate  shift (more results are presented in the appendix materials). We make the following observations. 
(i) Malware detectors encounter a significantly decreasing of accuracy and increasing of bNLL and bBSE with newer test data. This can be attributed to the natural software evolution that Google gradually updates Android APIs and practitioners upgrade their APKs to support new services. In particular, Droidetec suffer a lot from temporal covariate  shift and exhibits a low detection accuracy. As mentioned earlier, this may be that Droidetec is permitted to learn from a limited number of APIs, which inhibits handling the APKs with a broad range of APIs. 
(ii) Temp scaling has the same effect as the vanilla model in terms of bAccuracy; Ensemble enhances the vanilla models (DeepDrebin, MultimodalNN, and DeepDroid) at the start of several months but then this enhancement diminishes; VBI makes MultimodalNN to achieve the highest robustness under data evolution (but only achieves $\sim$80\% bAccuracy). 
(iii) Ensemble methods benefit calibration in terms of bNLL and bBSE when compared with the vanilla model; VBI incorporating DeepDrebin or MultimodalNN achieves an impressive result. 

Figure \ref{subfig:androzoo-ent-acc-bacc} plots the accuracy (at the upper half) and the balanced accuracy (at the lower half) after excluding the examples for which the detectors have entropy values greater than a threshold $\tau$. Figure \ref{subfig:androzoo-density} plots the sample density of predictive entropy.
We observe that 
(i) either accuracy or balanced accuracy decreases dramatically when the entropy increases, which is particularly true for DeepDroid and Droidetec. 
(ii) MultimodalNN incorporating VBI seems to work very well in terms of accuracy, but not necessary for balanced Accuracy. This is because the model classifies most benign samples correctly but not malicious ones until the threshold value is approach 0.9. Moreover, it is interesting to see that MultimodalNN incorporating VBI achieves the best bAccuracy without considering uncertainty (cf. Figure \ref{fig:androzoo-bacc-bnll-bbrier}, which is in sharp contrast to Figure \ref{subfig:androzoo-ent-acc-bacc}). The reason behind this is that MultimodalNN incorporating VBI correctly classifies a portion of examples with high entropy value, as shown in Figure \ref{subfig:androzoo-density}.
(iii) DeepDrebin incorporating VBI outperforms the other Drebin-based models, which resonates the results obtained in our second group of experiments (cf. Figure \ref{subfig:oos-02} in Section \ref{sec:aq2}). (iv) Figure \ref{subfig:androzoo-density} says that for all of the malware detectors except MultimodalNN incorporating VBI and DeepDrebin incorporating VBI, most examples tend to have a small entropy. This suggests ineffective calibrations of malware detectors under temporal covariate shifts and a lack of good calibration method.

\begin{insight}
	Calibrated malware detectors cannot cope with temporal datashit shift effectively, but VBI is promising for calibration and generalization under temporal covariate  shift.
\end{insight}

\subsection{Answering RQ4}

In order to quantify the predictive uncertainty of malware detectors under adversarial evasion attacks, we wage {\em transfer} attacks and generate adversarial APKs via a {\em surrogate} DeepDrebin model. We do not include adversarial APKs with respect to MultimodalNN, DeepDroid, and Droidetec because we do not find effective solutions. The surrogate DeepDrebin model consists of two fully-connected layers of 160 neurons with the ReLU activation function. We learn the model using the Adam optimizer with learning rate 0.001, batch size 128, and 150 epochs. We then generate adversarial examples against the surrogate model to perturb the 1,112 malicious APKs in the test dataset. Specifically, by following a recent study \cite{li2020adversarial}, we first perturb the feature vectors of the APKs using the {\em ``max'' PGDs+GDKDE} attack and the {\em Mimicry} attack, and then obtain adversarial APKs by using obfuscation techniques. In total, we obtain 1100 perturbed APK files for both attacks, respectively.

Table \ref{tab:drebin-adv-attack-full} summarizes the results of the malware detectors under the {\em ``max'' PGDs+GDKDE} attack and the Mimicry attack. Because the test dataset contains malware samples solely, we consider the Accuracy, NLL, BSE and ECE rather than their balanced versions. We observe that (i) the {\em ``max'' PGDs+GDKDE} attack renders DeepDrebin and MultimodalNN models useless, regardless of the calibration methods. Nevertheless, DeepDroid is robust against this attack because {\em opcode} are not used by the DeepDrebin and thus is unfocused by the attacker. However, DeepDroid still
suffers somewhat from this attack because of the {\em opcode} manipulations is leveraged for preserving the malicious functionality. 
(ii) Under the Mimicry attack, VBI makes DeepDrebin and MultimodalNN achieve the best accuracy and the lowest calibration error (in terms of NLL and BSE), while weighted Ensemble makes both obtain the worst results. However, this situation is changed in regards to DeepDroid and Droidetec. This might be that DeepDrebin and MultimodalNN are more sensitive to Mimicry attack than DeepDroid and Droidetec, leading to that an ensemble of vulnerable models decreases the robustness against the attack.

\begin{insight}
	Adversarial evasion attacks can render calibrated malware detectors, and therefore the quantified predictive uncertainty, useless, but heterogeneous feature extraction does improve the robustness of malware detectors against the transfer attacks.
\end{insight}

\section{Conclusion} \label{sec:conclusion}

We empirically quantified the predictive uncertainty of four deep malware detectors with six calibration strategies (i.e., 24 detectors in total). We found that the predictive uncertainty of calibrated malware detectors is useful except for adversarial examples. We hope this study will motivate and inspire more research in quantifying the uncertainty of malware detectors, which is of paramount importance in practice but it is currently little understood.

\begin{acks}
Q. Li is supported in part by the National Key R\&D Program of China under Grants 2020YFB1804604 and 2020YFB1804600, the 2020 Industrial Internet Innovation and Development Project from Ministry of Industry and Information Technology of China, the Fundamental Research Fund for the Central Universities under Grants 30918012204 and 30920041112, the 2019 Industrial Internet Innovation and Development Project from Ministry of Industry and Information Technology of China. S. Xu is supported in part by NSF Grants \#2122631 (\#1814825) and \#2115134, ARO Grant \#W911NF-17-1-0566, and Colorado State Bill 18-086.
\end{acks}


\bibliographystyle{ACM-Reference-Format}
\bibliography{uncertainty}


\begin{thebibliography}{50}


\ifx \showCODEN    \undefined \def \showCODEN     #1{\unskip}     \fi
\ifx \showDOI      \undefined \def \showDOI       #1{#1}\fi
\ifx \showISBNx    \undefined \def \showISBNx     #1{\unskip}     \fi
\ifx \showISBNxiii \undefined \def \showISBNxiii  #1{\unskip}     \fi
\ifx \showISSN     \undefined \def \showISSN      #1{\unskip}     \fi
\ifx \showLCCN     \undefined \def \showLCCN      #1{\unskip}     \fi
\ifx \shownote     \undefined \def \shownote      #1{#1}          \fi
\ifx \showarticletitle \undefined \def \showarticletitle #1{#1}   \fi
\ifx \showURL      \undefined \def \showURL       {\relax}        \fi
\providecommand\bibfield[2]{#2}
\providecommand\bibinfo[2]{#2}
\providecommand\natexlab[1]{#1}
\providecommand\showeprint[2][]{arXiv:#2}

\bibitem[\protect\citeauthoryear{Abadi, Agarwal, and et~al.}{Abadi
  et~al\mbox{.}}{2015}]%
        {dillon2017tensorflow}
\bibfield{author}{\bibinfo{person}{Mart\'{\i}n Abadi}, \bibinfo{person}{Ashish
  Agarwal}, {and} \bibinfo{person}{et al.}} \bibinfo{year}{2015}\natexlab{}.
\newblock \bibinfo{title}{{TensorFlow}: Large-Scale Machine Learning on
  Heterogeneous Systems}.
\newblock
\newblock
\urldef\tempurl%
\url{https://www.tensorflow.org/}
\showURL{%
\tempurl}
\newblock
\shownote{Software available from tensorflow.org.}


\bibitem[\protect\citeauthoryear{Abadi, Barham, et~al\mbox{.}}{Abadi
  et~al\mbox{.}}{2016}]%
        {abadi2016tensorflow}
\bibfield{author}{\bibinfo{person}{Mart{\'\i}n Abadi}, \bibinfo{person}{Paul
  Barham}, {et~al\mbox{.}}} \bibinfo{year}{2016}\natexlab{}.
\newblock \showarticletitle{Tensorflow: A system for large-scale machine
  learning}. In \bibinfo{booktitle}{\emph{OSDI' 16}}.
  \bibinfo{publisher}{{USENIX} Association}, \bibinfo{address}{Savannah, GA,
  USA}, \bibinfo{pages}{265--283}.
\newblock


\bibitem[\protect\citeauthoryear{Allix, Bissyand{\'e}, et~al\mbox{.}}{Allix
  et~al\mbox{.}}{2016}]%
        {Allix:2016:ACM:2901739.2903508}
\bibfield{author}{\bibinfo{person}{Kevin Allix},
  \bibinfo{person}{Tegawend{\'e}~F. Bissyand{\'e}}, {et~al\mbox{.}}}
  \bibinfo{year}{2016}\natexlab{}.
\newblock \showarticletitle{AndroZoo: Collecting Millions of Android Apps for
  the Research Community}. In \bibinfo{booktitle}{\emph{International
  Conference on MSR}} (Austin, Texas). \bibinfo{publisher}{ACM},
  \bibinfo{address}{NY, USA}, \bibinfo{pages}{468--471}.
\newblock
\showISBNx{978-1-4503-4186-8}
\urldef\tempurl%
\url{https://doi.org/10.1145/2901739.2903508}
\showDOI{\tempurl}


\bibitem[\protect\citeauthoryear{Arp, Spreitzenbarth, et~al\mbox{.}}{Arp
  et~al\mbox{.}}{2014}]%
        {arp2014drebin}
\bibfield{author}{\bibinfo{person}{Daniel Arp}, \bibinfo{person}{Michael
  Spreitzenbarth}, {et~al\mbox{.}}} \bibinfo{year}{2014}\natexlab{}.
\newblock \showarticletitle{Drebin: Effective and explainable detection of
  android malware in your pocket}. In \bibinfo{booktitle}{\emph{NDSS}},
  Vol.~\bibinfo{volume}{14}. \bibinfo{publisher}{The Internet Society},
  \bibinfo{address}{San Diego, California, USA}, \bibinfo{pages}{23--26}.
\newblock


\bibitem[\protect\citeauthoryear{Bhatt, Zhang, and et~al}{Bhatt
  et~al\mbox{.}}{2020}]%
        {DBLP:journals/corr/abs-2011-07586}
\bibfield{author}{\bibinfo{person}{Umang Bhatt}, \bibinfo{person}{Yunfeng
  Zhang}, {and} \bibinfo{person}{et al}.} \bibinfo{year}{2020}\natexlab{}.
\newblock \showarticletitle{Uncertainty as a Form of Transparency: Measuring,
  Communicating, and Using Uncertainty}.
\newblock \bibinfo{journal}{\emph{CoRR}}  \bibinfo{volume}{abs/2011.07586}
  (\bibinfo{year}{2020}).
\newblock
\urldef\tempurl%
\url{https://arxiv.org/abs/2011.07586}
\showURL{%
\tempurl}


\bibitem[\protect\citeauthoryear{Blundell, Cornebise, Kavukcuoglu, and
  Wierstra}{Blundell et~al\mbox{.}}{2015}]%
        {blundell2015weight}
\bibfield{author}{\bibinfo{person}{Charles Blundell}, \bibinfo{person}{Julien
  Cornebise}, \bibinfo{person}{Koray Kavukcuoglu}, {and} \bibinfo{person}{Daan
  Wierstra}.} \bibinfo{year}{2015}\natexlab{}.
\newblock \showarticletitle{Weight Uncertainty in Neural Network}. In
  \bibinfo{booktitle}{\emph{Proceedings of the 32nd International Conference on
  Machine Learning}}, Vol.~\bibinfo{volume}{37}. \bibinfo{publisher}{PMLR},
  \bibinfo{address}{Lille, France}, \bibinfo{pages}{1613--1622}.
\newblock
\urldef\tempurl%
\url{https://proceedings.mlr.press/v37/blundell15.html}
\showURL{%
\tempurl}


\bibitem[\protect\citeauthoryear{Brier}{Brier}{1950}]%
        {brier1950verification}
\bibfield{author}{\bibinfo{person}{Glenn~W Brier}.}
  \bibinfo{year}{1950}\natexlab{}.
\newblock \showarticletitle{Verification of forecasts expressed in terms of
  probability}.
\newblock \bibinfo{journal}{\emph{Monthly weather review}}
  \bibinfo{volume}{78}, \bibinfo{number}{1} (\bibinfo{year}{1950}),
  \bibinfo{pages}{1--3}.
\newblock


\bibitem[\protect\citeauthoryear{{Brodersen}, {Ong}, {Stephan}, and
  {Buhmann}}{{Brodersen} et~al\mbox{.}}{2010}]%
        {5597285}
\bibfield{author}{\bibinfo{person}{K.~H. {Brodersen}}, \bibinfo{person}{C.~S.
  {Ong}}, \bibinfo{person}{K.~E. {Stephan}}, {and} \bibinfo{person}{J.~M.
  {Buhmann}}.} \bibinfo{year}{2010}\natexlab{}.
\newblock \showarticletitle{The Balanced Accuracy and Its Posterior
  Distribution}. In \bibinfo{booktitle}{\emph{2010 20th International
  Conference on Pattern Recognition}}. \bibinfo{publisher}{{IEEE} Computer
  Society}, \bibinfo{address}{Istanbul, Turkey}, \bibinfo{pages}{3121--3124}.
\newblock


\bibitem[\protect\citeauthoryear{Candela, Rasmussen, Sinz, Bousquet, and
  Sch{\"{o}}lkopf}{Candela et~al\mbox{.}}{2005}]%
        {DBLP:conf/mlcw/CandelaRSBS05}
\bibfield{author}{\bibinfo{person}{Joaquin~Qui{\~{n}}onero Candela},
  \bibinfo{person}{Carl~Edward Rasmussen}, \bibinfo{person}{Fabian~H. Sinz},
  \bibinfo{person}{Olivier Bousquet}, {and} \bibinfo{person}{Bernhard
  Sch{\"{o}}lkopf}.} \bibinfo{year}{2005}\natexlab{}.
\newblock \showarticletitle{Evaluating Predictive Uncertainty Challenge}. In
  \bibinfo{booktitle}{\emph{Machine Learning Challenges, Evaluating Predictive
  Uncertainty, Visual Object Classification and Recognizing Textual Entailment,
  First {PASCAL} Machine Learning Challenges Workshop}},
  Vol.~\bibinfo{volume}{3944}. \bibinfo{publisher}{Springer},
  \bibinfo{address}{Southampton, UK}, \bibinfo{pages}{1--27}.
\newblock


\bibitem[\protect\citeauthoryear{Chen, Ye, and Bourlai}{Chen
  et~al\mbox{.}}{2017}]%
        {DBLP:conf/eisic/ChenYB17}
\bibfield{author}{\bibinfo{person}{Lingwei Chen}, \bibinfo{person}{Yanfang Ye},
  {and} \bibinfo{person}{Thirimachos Bourlai}.}
  \bibinfo{year}{2017}\natexlab{}.
\newblock \showarticletitle{Adversarial Machine Learning in Malware Detection:
  Arms Race between Evasion Attack and Defense}. In
  \bibinfo{booktitle}{\emph{EISIC'2017}}. \bibinfo{publisher}{{IEEE} Computer
  Society}, \bibinfo{address}{Athens, Greece}, \bibinfo{pages}{99--106}.
\newblock


\bibitem[\protect\citeauthoryear{Corvus}{Corvus}{2020}]%
        {VirusShare:Online}
\bibfield{author}{\bibinfo{person}{Forensics Corvus}.}
  \bibinfo{year}{2020}\natexlab{}.
\newblock \bibinfo{booktitle}{\emph{VirusShare}}.
\newblock
\urldef\tempurl%
\url{https://virusshare.com/}
\showURL{%
\tempurl}


\bibitem[\protect\citeauthoryear{Demontis, Melis, et~al\mbox{.}}{Demontis
  et~al\mbox{.}}{2017}]%
        {demontis2017yes}
\bibfield{author}{\bibinfo{person}{Ambra Demontis}, \bibinfo{person}{Marco
  Melis}, {et~al\mbox{.}}} \bibinfo{year}{2017}\natexlab{}.
\newblock \showarticletitle{Yes, machine learning can be more secure! a case
  study on android malware detection}.
\newblock \bibinfo{journal}{\emph{IEEE TDSC}} \bibinfo{volume}{16},
  \bibinfo{number}{4} (\bibinfo{year}{2017}), \bibinfo{pages}{711--724}.
\newblock


\bibitem[\protect\citeauthoryear{Desnos}{Desnos}{2020}]%
        {Androguard:Online}
\bibfield{author}{\bibinfo{person}{Anthony Desnos}.}
  \bibinfo{year}{2020}\natexlab{}.
\newblock \bibinfo{booktitle}{\emph{Androguard}}.
\newblock
\urldef\tempurl%
\url{https://github.com/androguard/androguard}
\showURL{%
\tempurl}


\bibitem[\protect\citeauthoryear{Du, Sun, Chen, Cho, and Xu}{Du
  et~al\mbox{.}}{2018}]%
        {DBLP:journals/tifs/DuSCCX18}
\bibfield{author}{\bibinfo{person}{Pang Du}, \bibinfo{person}{Zheyuan Sun},
  \bibinfo{person}{Huashan Chen}, \bibinfo{person}{Jin{-}Hee Cho}, {and}
  \bibinfo{person}{Shouhuai Xu}.} \bibinfo{year}{2018}\natexlab{}.
\newblock \showarticletitle{Statistical Estimation of Malware Detection Metrics
  in the Absence of Ground Truth}.
\newblock \bibinfo{journal}{\emph{{IEEE} Trans. Inf. Forensics Secur.}}
  \bibinfo{volume}{13}, \bibinfo{number}{12} (\bibinfo{year}{2018}),
  \bibinfo{pages}{2965--2980}.
\newblock


\bibitem[\protect\citeauthoryear{Gal and Ghahramani}{Gal and
  Ghahramani}{2016}]%
        {gal2016dropout}
\bibfield{author}{\bibinfo{person}{Yarin Gal} {and} \bibinfo{person}{Zoubin
  Ghahramani}.} \bibinfo{year}{2016}\natexlab{}.
\newblock \showarticletitle{Dropout as a bayesian approximation: Representing
  model uncertainty in deep learning}. In
  \bibinfo{booktitle}{\emph{international conference on machine learning}}.
  \bibinfo{publisher}{JMLR.org}, \bibinfo{address}{NY, USA},
  \bibinfo{pages}{1050--1059}.
\newblock


\bibitem[\protect\citeauthoryear{Graves}{Graves}{2011}]%
        {DBLP:conf/nips/Graves11}
\bibfield{author}{\bibinfo{person}{Alex Graves}.}
  \bibinfo{year}{2011}\natexlab{}.
\newblock \showarticletitle{Practical Variational Inference for Neural
  Networks}. In \bibinfo{booktitle}{\emph{Advances in Neural Information
  Processing Systems 24: 25th Annual Conference on Neural Information
  Processing Systems 2011}}. \bibinfo{publisher}{Curran Associates Inc.},
  \bibinfo{address}{Granada, Spain}, \bibinfo{pages}{2348--2356}.
\newblock


\bibitem[\protect\citeauthoryear{Grosse, Papernot, et~al\mbox{.}}{Grosse
  et~al\mbox{.}}{2017}]%
        {grosse2017adversarial}
\bibfield{author}{\bibinfo{person}{Kathrin Grosse}, \bibinfo{person}{Nicolas
  Papernot}, {et~al\mbox{.}}} \bibinfo{year}{2017}\natexlab{}.
\newblock \showarticletitle{Adversarial examples for malware detection}. In
  \bibinfo{booktitle}{\emph{ESORICS}}. \bibinfo{publisher}{Springer},
  \bibinfo{address}{Oslo, Norway}, \bibinfo{pages}{62--79}.
\newblock


\bibitem[\protect\citeauthoryear{Guo, Pleiss, Sun, and Weinberger}{Guo
  et~al\mbox{.}}{2017}]%
        {guo2017calibration}
\bibfield{author}{\bibinfo{person}{Chuan Guo}, \bibinfo{person}{Geoff Pleiss},
  \bibinfo{person}{Yu Sun}, {and} \bibinfo{person}{Kilian~Q. Weinberger}.}
  \bibinfo{year}{2017}\natexlab{}.
\newblock \showarticletitle{On Calibration of Modern Neural Networks}. In
  \bibinfo{booktitle}{\emph{Proceedings of the 34th International Conference on
  Machine Learning, {ICML}}} \emph{(\bibinfo{series}{Proceedings of Machine
  Learning Research}, Vol.~\bibinfo{volume}{70})},
  \bibfield{editor}{\bibinfo{person}{Doina Precup} {and}
  \bibinfo{person}{Yee~Whye Teh}} (Eds.). \bibinfo{publisher}{{PMLR}},
  \bibinfo{address}{Sydney, NSW, Australia}, \bibinfo{pages}{1321--1330}.
\newblock


\bibitem[\protect\citeauthoryear{{Huang} and {Kao}}{{Huang} and {Kao}}{2018}]%
        {8622324}
\bibfield{author}{\bibinfo{person}{T.~H. {Huang}} {and} \bibinfo{person}{H.
  {Kao}}.} \bibinfo{year}{2018}\natexlab{}.
\newblock \showarticletitle{R2-D2: ColoR-inspired Convolutional NeuRal Network
  (CNN)-based AndroiD Malware Detections}. In \bibinfo{booktitle}{\emph{2018
  IEEE International Conference on Big Data (Big Data)}}.
  \bibinfo{publisher}{{IEEE}}, \bibinfo{address}{Seattle, WA, USA},
  \bibinfo{pages}{2633--2642}.
\newblock


\bibitem[\protect\citeauthoryear{Jordaney, Sharad, et~al\mbox{.}}{Jordaney
  et~al\mbox{.}}{2017}]%
        {203684}
\bibfield{author}{\bibinfo{person}{Roberto Jordaney}, \bibinfo{person}{Kumar
  Sharad}, {et~al\mbox{.}}} \bibinfo{year}{2017}\natexlab{}.
\newblock \showarticletitle{Transcend: Detecting Concept Drift in Malware
  Classification Models}. In \bibinfo{booktitle}{\emph{{USENIX} Security 17}}.
  \bibinfo{publisher}{{USENIX} Association}, \bibinfo{address}{Vancouver, BC},
  \bibinfo{pages}{625--642}.
\newblock
\showISBNx{978-1-931971-40-9}
\urldef\tempurl%
\url{https://www.usenix.org/conference/usenixsecurity17/technical-sessions/presentation/jordaney}
\showURL{%
\tempurl}


\bibitem[\protect\citeauthoryear{Kendall and Gal}{Kendall and Gal}{2017}]%
        {kendall2017uncertainties}
\bibfield{author}{\bibinfo{person}{Alex Kendall} {and} \bibinfo{person}{Yarin
  Gal}.} \bibinfo{year}{2017}\natexlab{}.
\newblock \showarticletitle{What uncertainties do we need in bayesian deep
  learning for computer vision?}. In \bibinfo{booktitle}{\emph{NeurIPS}}.
  \bibinfo{publisher}{Curran Associates Inc.}, \bibinfo{address}{Long Beach,
  CA, {USA}}, \bibinfo{pages}{5574--5584}.
\newblock


\bibitem[\protect\citeauthoryear{{Kim}, {Kang}, et~al\mbox{.}}{{Kim}
  et~al\mbox{.}}{2019}]%
        {8443370}
\bibfield{author}{\bibinfo{person}{T. {Kim}}, \bibinfo{person}{B. {Kang}},
  {et~al\mbox{.}}} \bibinfo{year}{2019}\natexlab{}.
\newblock \showarticletitle{A Multimodal Deep Learning Method for Android
  Malware Detection Using Various Features}.
\newblock \bibinfo{journal}{\emph{IEEE Trans. Info. Forensics and Sec.}}
  \bibinfo{volume}{14}, \bibinfo{number}{3} (\bibinfo{year}{2019}),
  \bibinfo{pages}{773--788}.
\newblock


\bibitem[\protect\citeauthoryear{Kim}{Kim}{2014}]%
        {kim2014convolutional}
\bibfield{author}{\bibinfo{person}{Yoon Kim}.} \bibinfo{year}{2014}\natexlab{}.
\newblock \showarticletitle{Convolutional Neural Networks for Sentence
  Classification}. In \bibinfo{booktitle}{\emph{Proceedings of the 2014
  Conference on Empirical Methods in Natural Language Processing, {EMNLP}}},
  \bibfield{editor}{\bibinfo{person}{Alessandro Moschitti},
  \bibinfo{person}{Bo~Pang}, {and} \bibinfo{person}{Walter Daelemans}} (Eds.).
  \bibinfo{publisher}{{ACL}}, \bibinfo{address}{Doha, Qatar},
  \bibinfo{pages}{1746--1751}.
\newblock


\bibitem[\protect\citeauthoryear{Lab}{Lab}{2020}]%
        {kaspersky:Online}
\bibfield{author}{\bibinfo{person}{Kaspersky Lab}.}
  \bibinfo{year}{2020}\natexlab{}.
\newblock \bibinfo{booktitle}{\emph{Kaspersky}}.
\newblock
\urldef\tempurl%
\url{https://www.kaspersky.com}
\showURL{%
\tempurl}


\bibitem[\protect\citeauthoryear{Lakshminarayanan, Pritzel, and
  Blundell}{Lakshminarayanan et~al\mbox{.}}{2017}]%
        {lakshminarayanan2017simple}
\bibfield{author}{\bibinfo{person}{Balaji Lakshminarayanan},
  \bibinfo{person}{Alexander Pritzel}, {and} \bibinfo{person}{Charles
  Blundell}.} \bibinfo{year}{2017}\natexlab{}.
\newblock \showarticletitle{Simple and scalable predictive uncertainty
  estimation using deep ensembles}. In \bibinfo{booktitle}{\emph{NeurIPS}}.
  \bibinfo{publisher}{Curran Associates Inc.}, \bibinfo{address}{Long Beach,
  CA, {USA}}, \bibinfo{pages}{6402--6413}.
\newblock


\bibitem[\protect\citeauthoryear{Leibig, Allken, Ayhan, Berens, and
  Wahl}{Leibig et~al\mbox{.}}{2017}]%
        {leibig2017leveraging}
\bibfield{author}{\bibinfo{person}{Christian Leibig}, \bibinfo{person}{Vaneeda
  Allken}, \bibinfo{person}{Murat~Se{\c{c}}kin Ayhan}, \bibinfo{person}{Philipp
  Berens}, {and} \bibinfo{person}{Siegfried Wahl}.}
  \bibinfo{year}{2017}\natexlab{}.
\newblock \showarticletitle{Leveraging uncertainty information from deep neural
  networks for disease detection}.
\newblock \bibinfo{journal}{\emph{Scientific reports}} \bibinfo{volume}{7},
  \bibinfo{number}{1} (\bibinfo{year}{2017}), \bibinfo{pages}{1--14}.
\newblock


\bibitem[\protect\citeauthoryear{Li and Li}{Li and Li}{2020}]%
        {li2020adversarial}
\bibfield{author}{\bibinfo{person}{Deqiang Li} {and} \bibinfo{person}{Qianmu
  Li}.} \bibinfo{year}{2020}\natexlab{}.
\newblock \showarticletitle{Adversarial Deep Ensemble: Evasion Attacks and
  Defenses for Malware Detection}.
\newblock \bibinfo{journal}{\emph{IEEE Trans. Info. Forensics and Sec.}}
  \bibinfo{volume}{15} (\bibinfo{year}{2020}), \bibinfo{pages}{3886--3900}.
\newblock


\bibitem[\protect\citeauthoryear{Lu, Liu, Dong, Gu, Gama, and Zhang}{Lu
  et~al\mbox{.}}{2019}]%
        {DBLP:journals/tkde/LuLDGGZ19}
\bibfield{author}{\bibinfo{person}{Jie Lu}, \bibinfo{person}{Anjin Liu},
  \bibinfo{person}{Fan Dong}, \bibinfo{person}{Feng Gu},
  \bibinfo{person}{Jo{\~{a}}o Gama}, {and} \bibinfo{person}{Guangquan Zhang}.}
  \bibinfo{year}{2019}\natexlab{}.
\newblock \showarticletitle{Learning under Concept Drift: {A} Review}.
\newblock \bibinfo{journal}{\emph{{IEEE} Trans. Knowl. Data Eng.}}
  \bibinfo{volume}{31}, \bibinfo{number}{12} (\bibinfo{year}{2019}),
  \bibinfo{pages}{2346--2363}.
\newblock


\bibitem[\protect\citeauthoryear{Ma, Ge, Wang, Liu, and Liu}{Ma
  et~al\mbox{.}}{2020}]%
        {ma2020droidetec}
\bibfield{author}{\bibinfo{person}{Zhuo Ma}, \bibinfo{person}{Haoran Ge},
  \bibinfo{person}{Zhuzhu Wang}, \bibinfo{person}{Yang Liu}, {and}
  \bibinfo{person}{Ximeng Liu}.} \bibinfo{year}{2020}\natexlab{}.
\newblock \showarticletitle{Droidetec: Android malware detection and malicious
  code localization through deep learning}.
\newblock \bibinfo{journal}{\emph{CoRR}}  \bibinfo{volume}{abs/2002.03594}
  (\bibinfo{year}{2020}).
\newblock
\urldef\tempurl%
\url{https://arxiv.org/abs/2002.03594}
\showURL{%
\tempurl}


\bibitem[\protect\citeauthoryear{McLaughlin, Martinez~del Rincon,
  et~al\mbox{.}}{McLaughlin et~al\mbox{.}}{2017}]%
        {10.1145/3029806.3029823}
\bibfield{author}{\bibinfo{person}{Niall McLaughlin}, \bibinfo{person}{Jesus
  Martinez~del Rincon}, {et~al\mbox{.}}} \bibinfo{year}{2017}\natexlab{}.
\newblock \showarticletitle{Deep Android Malware Detection}. In
  \bibinfo{booktitle}{\emph{CODASPY ’17}} (Scottsdale, Arizona, USA).
  \bibinfo{publisher}{ACM}, \bibinfo{address}{NY, USA},
  \bibinfo{pages}{301–308}.
\newblock
\showISBNx{9781450345231}
\urldef\tempurl%
\url{https://doi.org/10.1145/3029806.3029823}
\showDOI{\tempurl}


\bibitem[\protect\citeauthoryear{Naeini, Cooper, and Hauskrecht}{Naeini
  et~al\mbox{.}}{2015}]%
        {naeini2015obtaining}
\bibfield{author}{\bibinfo{person}{Mahdi~Pakdaman Naeini},
  \bibinfo{person}{Gregory Cooper}, {and} \bibinfo{person}{Milos Hauskrecht}.}
  \bibinfo{year}{2015}\natexlab{}.
\newblock \showarticletitle{Obtaining well calibrated probabilities using
  bayesian binning}. In \bibinfo{booktitle}{\emph{Twenty-Ninth AAAI Conference
  on Artificial Intelligence}}. \bibinfo{publisher}{{AAAI} Press},
  \bibinfo{address}{Austin, Texas, {USA}}, \bibinfo{pages}{2901--2907}.
\newblock


\bibitem[\protect\citeauthoryear{Nguyen, Raff, Nicholas, and Holt}{Nguyen
  et~al\mbox{.}}{2021}]%
        {DBLP:journals/corr/abs-2108-04081}
\bibfield{author}{\bibinfo{person}{Andr{\'{e}}~T. Nguyen},
  \bibinfo{person}{Edward Raff}, \bibinfo{person}{Charles Nicholas}, {and}
  \bibinfo{person}{James Holt}.} \bibinfo{year}{2021}\natexlab{}.
\newblock \showarticletitle{Leveraging Uncertainty for Improved Static Malware
  Detection Under Extreme False Positive Constraints}.
\newblock \bibinfo{journal}{\emph{CoRR}}  \bibinfo{volume}{abs/2108.04081}
  (\bibinfo{year}{2021}).
\newblock
\urldef\tempurl%
\url{https://arxiv.org/abs/2108.04081}
\showURL{%
\tempurl}


\bibitem[\protect\citeauthoryear{Niculescu-Mizil and Caruana}{Niculescu-Mizil
  and Caruana}{2005}]%
        {niculescu2005predicting}
\bibfield{author}{\bibinfo{person}{Alexandru Niculescu-Mizil} {and}
  \bibinfo{person}{Rich Caruana}.} \bibinfo{year}{2005}\natexlab{}.
\newblock \showarticletitle{Predicting good probabilities with supervised
  learning}. In \bibinfo{booktitle}{\emph{ICML}}. \bibinfo{publisher}{{ACM}},
  \bibinfo{address}{Bonn, Germany}, \bibinfo{pages}{625--632}.
\newblock


\bibitem[\protect\citeauthoryear{Pearce, Leibfried, et~al\mbox{.}}{Pearce
  et~al\mbox{.}}{2020}]%
        {pearce2020uncertainty}
\bibfield{author}{\bibinfo{person}{Tim Pearce}, \bibinfo{person}{Felix
  Leibfried}, {et~al\mbox{.}}} \bibinfo{year}{2020}\natexlab{}.
\newblock \showarticletitle{Uncertainty in Neural Networks: Approximately
  Bayesian Ensembling}. In \bibinfo{booktitle}{\emph{AISTATS}}.
  \bibinfo{publisher}{{PMLR}}, \bibinfo{address}{Online [Palermo, Sicily,
  Italy]}, \bibinfo{pages}{234--244}.
\newblock


\bibitem[\protect\citeauthoryear{Pendlebury, Pierazzi,
  et~al\mbox{.}}{Pendlebury et~al\mbox{.}}{2019}]%
        {235493}
\bibfield{author}{\bibinfo{person}{Feargus Pendlebury}, \bibinfo{person}{Fabio
  Pierazzi}, {et~al\mbox{.}}} \bibinfo{year}{2019}\natexlab{}.
\newblock \showarticletitle{{TESSERACT}: Eliminating Experimental Bias in
  Malware Classification across Space and Time}. In
  \bibinfo{booktitle}{\emph{{USENIX} Security 19}}.
  \bibinfo{publisher}{{USENIX} Association}, \bibinfo{address}{Santa Clara,
  CA}, \bibinfo{pages}{729--746}.
\newblock
\showISBNx{978-1-939133-06-9}
\urldef\tempurl%
\url{https://www.usenix.org/conference/usenixsecurity19/presentation/pendlebury}
\showURL{%
\tempurl}


\bibitem[\protect\citeauthoryear{Pendleton, Garcia-Lebron, Cho, and
  Xu}{Pendleton et~al\mbox{.}}{2016}]%
        {Pendleton:2016}
\bibfield{author}{\bibinfo{person}{Marcus Pendleton}, \bibinfo{person}{Richard
  Garcia-Lebron}, \bibinfo{person}{Jin-Hee Cho}, {and}
  \bibinfo{person}{Shouhuai Xu}.} \bibinfo{year}{2016}\natexlab{}.
\newblock \showarticletitle{A Survey on Systems Security Metrics}.
\newblock \bibinfo{journal}{\emph{ACM Comput. Surv.}} \bibinfo{volume}{49},
  \bibinfo{number}{4} (\bibinfo{date}{Dec.} \bibinfo{year}{2016}),
  \bibinfo{pages}{1--35}.
\newblock


\bibitem[\protect\citeauthoryear{Quionero-Candela, Sugiyama, Schwaighofer, and
  Lawrence}{Quionero-Candela et~al\mbox{.}}{2009}]%
        {10.5555/1462129}
\bibfield{author}{\bibinfo{person}{Joaquin Quionero-Candela},
  \bibinfo{person}{Masashi Sugiyama}, \bibinfo{person}{Anton Schwaighofer},
  {and} \bibinfo{person}{Neil~D. Lawrence}.} \bibinfo{year}{2009}\natexlab{}.
\newblock \bibinfo{booktitle}{\emph{Dataset Shift in Machine Learning}}.
\newblock \bibinfo{publisher}{The MIT Press}, \bibinfo{address}{Cambridge, MA}.
\newblock
\showISBNx{0262170051}


\bibitem[\protect\citeauthoryear{Sistemas}{Sistemas}{2020}]%
        {VirusTotal:Online}
\bibfield{author}{\bibinfo{person}{Hispasec Sistemas}.}
  \bibinfo{year}{2020}\natexlab{}.
\newblock \bibinfo{booktitle}{\emph{VirusTotal}}.
\newblock Alphabet, Inc.
\newblock
\urldef\tempurl%
\url{https://www.virustotal.com}
\showURL{%
\tempurl}


\bibitem[\protect\citeauthoryear{Snoek, Ovadia, Fertig, Lakshminarayanan,
  Nowozin, Sculley, Dillon, Ren, and Nado}{Snoek et~al\mbox{.}}{2019}]%
        {snoek2019can}
\bibfield{author}{\bibinfo{person}{Jasper Snoek}, \bibinfo{person}{Yaniv
  Ovadia}, \bibinfo{person}{Emily Fertig}, \bibinfo{person}{Balaji
  Lakshminarayanan}, \bibinfo{person}{Sebastian Nowozin}, \bibinfo{person}{D
  Sculley}, \bibinfo{person}{Joshua Dillon}, \bibinfo{person}{Jie Ren}, {and}
  \bibinfo{person}{Zachary Nado}.} \bibinfo{year}{2019}\natexlab{}.
\newblock \showarticletitle{Can you trust your model's uncertainty? Evaluating
  predictive uncertainty under dataset shift}. In
  \bibinfo{booktitle}{\emph{Advances in Neural Information Processing
  Systems}}. \bibinfo{publisher}{Curran Associates Inc.},
  \bibinfo{address}{Vancouver, BC, Canada}, \bibinfo{pages}{13969--13980}.
\newblock


\bibitem[\protect\citeauthoryear{Srivastava, Hinton, Krizhevsky, Sutskever, and
  Salakhutdinov}{Srivastava et~al\mbox{.}}{2014}]%
        {srivastava2014dropout}
\bibfield{author}{\bibinfo{person}{Nitish Srivastava},
  \bibinfo{person}{Geoffrey Hinton}, \bibinfo{person}{Alex Krizhevsky},
  \bibinfo{person}{Ilya Sutskever}, {and} \bibinfo{person}{Ruslan
  Salakhutdinov}.} \bibinfo{year}{2014}\natexlab{}.
\newblock \showarticletitle{Dropout: a simple way to prevent neural networks
  from overfitting}.
\newblock \bibinfo{journal}{\emph{The journal of machine learning research}}
  \bibinfo{volume}{15}, \bibinfo{number}{1} (\bibinfo{year}{2014}),
  \bibinfo{pages}{1929--1958}.
\newblock


\bibitem[\protect\citeauthoryear{Thireou and Reczko}{Thireou and
  Reczko}{2007}]%
        {thireou2007bidirectional}
\bibfield{author}{\bibinfo{person}{Trias Thireou} {and} \bibinfo{person}{Martin
  Reczko}.} \bibinfo{year}{2007}\natexlab{}.
\newblock \showarticletitle{Bidirectional long short-term memory networks for
  predicting the subcellular localization of eukaryotic proteins}.
\newblock \bibinfo{journal}{\emph{IEEE/ACM transactions on computational
  biology and bioinformatics}} \bibinfo{volume}{4}, \bibinfo{number}{3}
  (\bibinfo{year}{2007}), \bibinfo{pages}{441--446}.
\newblock


\bibitem[\protect\citeauthoryear{Tumbleson and Wi\'sniewski}{Tumbleson and
  Wi\'sniewski}{2020}]%
        {apktool:Online}
\bibfield{author}{\bibinfo{person}{Connor Tumbleson} {and}
  \bibinfo{person}{Ryszard Wi\'sniewski}.} \bibinfo{year}{2020}\natexlab{}.
\newblock \bibinfo{booktitle}{\emph{Apktool}}.
\newblock
\urldef\tempurl%
\url{https://ibotpeaches.github.io/Apktool}
\showURL{%
\tempurl}


\bibitem[\protect\citeauthoryear{Vaicenavicius, Widmann, Andersson, Lindsten,
  Roll, and Sch{\"{o}}n}{Vaicenavicius et~al\mbox{.}}{2019}]%
        {vaicenavicius2019evaluating}
\bibfield{author}{\bibinfo{person}{Juozas Vaicenavicius},
  \bibinfo{person}{David Widmann}, \bibinfo{person}{Carl~R. Andersson},
  \bibinfo{person}{Fredrik Lindsten}, \bibinfo{person}{Jacob Roll}, {and}
  \bibinfo{person}{Thomas~B. Sch{\"{o}}n}.} \bibinfo{year}{2019}\natexlab{}.
\newblock \showarticletitle{Evaluating model calibration in classification}. In
  \bibinfo{booktitle}{\emph{The 22nd International Conference on Artificial
  Intelligence and Statistics, {AISTATS}}}, Vol.~\bibinfo{volume}{89}.
  \bibinfo{publisher}{{PMLR}}, \bibinfo{address}{Naha, Okinawa, Japan},
  \bibinfo{pages}{3459--3467}.
\newblock


\bibitem[\protect\citeauthoryear{Wang, Lawrence, and Niepert}{Wang
  et~al\mbox{.}}{2021}]%
        {DBLP:conf/iclr/WangLN21}
\bibfield{author}{\bibinfo{person}{Cheng Wang}, \bibinfo{person}{Carolin
  Lawrence}, {and} \bibinfo{person}{Mathias Niepert}.}
  \bibinfo{year}{2021}\natexlab{}.
\newblock \showarticletitle{Uncertainty Estimation and Calibration with
  Finite-State Probabilistic RNNs}. In \bibinfo{booktitle}{\emph{9th
  International Conference on Learning Representations}}.
  \bibinfo{publisher}{OpenReview.net}, \bibinfo{address}{Virtual Event,
  Austria}.
\newblock


\bibitem[\protect\citeauthoryear{Welling and Teh}{Welling and Teh}{2011}]%
        {welling2011bayesian}
\bibfield{author}{\bibinfo{person}{Max Welling} {and} \bibinfo{person}{Yee~W
  Teh}.} \bibinfo{year}{2011}\natexlab{}.
\newblock \showarticletitle{Bayesian learning via stochastic gradient Langevin
  dynamics}. In \bibinfo{booktitle}{\emph{ICML}}.
  \bibinfo{publisher}{Omnipress}, \bibinfo{address}{Madison, WI, USA},
  \bibinfo{pages}{681--688}.
\newblock


\bibitem[\protect\citeauthoryear{Widmer and Kubat}{Widmer and Kubat}{1996}]%
        {DBLP:journals/ml/WidmerK96}
\bibfield{author}{\bibinfo{person}{Gerhard Widmer} {and}
  \bibinfo{person}{Miroslav Kubat}.} \bibinfo{year}{1996}\natexlab{}.
\newblock \showarticletitle{Learning in the Presence of Concept Drift and
  Hidden Contexts}.
\newblock \bibinfo{journal}{\emph{Mach. Learn.}} \bibinfo{volume}{23},
  \bibinfo{number}{1} (\bibinfo{year}{1996}), \bibinfo{pages}{69--101}.
\newblock


\bibitem[\protect\citeauthoryear{Xu, Zhan, Xu, and Ye}{Xu
  et~al\mbox{.}}{2014}]%
        {DBLP:conf/cns/XuZXY14}
\bibfield{author}{\bibinfo{person}{Li Xu}, \bibinfo{person}{Zhenxin Zhan},
  \bibinfo{person}{Shouhuai Xu}, {and} \bibinfo{person}{Keying Ye}.}
  \bibinfo{year}{2014}\natexlab{}.
\newblock \showarticletitle{An evasion and counter-evasion study in malicious
  websites detection}. In \bibinfo{booktitle}{\emph{{IEEE} Conference on
  Communications and Network Security (CNS'2014)}}.
  \bibinfo{publisher}{{IEEE}}, \bibinfo{address}{San Francisco, CA, USA},
  \bibinfo{pages}{265--273}.
\newblock


\bibitem[\protect\citeauthoryear{Ye, Li, and et~al.}{Ye et~al\mbox{.}}{2017}]%
        {DBLP:journals/csur/YeLAI17}
\bibfield{author}{\bibinfo{person}{Yanfang Ye}, \bibinfo{person}{Tao Li}, {and}
  \bibinfo{person}{et al.}} \bibinfo{year}{2017}\natexlab{}.
\newblock \showarticletitle{A Survey on Malware Detection Using Data Mining
  Techniques}.
\newblock \bibinfo{journal}{\emph{{ACM} Comput. Surv.}} \bibinfo{volume}{50},
  \bibinfo{number}{3} (\bibinfo{year}{2017}), \bibinfo{pages}{41:1--41:40}.
\newblock


\bibitem[\protect\citeauthoryear{Zhang, Zhang, and et~al.}{Zhang
  et~al\mbox{.}}{2020}]%
        {10.1145-3372297.3417291}
\bibfield{author}{\bibinfo{person}{Xiaohan Zhang}, \bibinfo{person}{Yuan
  Zhang}, {and} \bibinfo{person}{et al.}} \bibinfo{year}{2020}\natexlab{}.
\newblock \showarticletitle{Enhancing State-of-the-Art Classifiers with API
  Semantics to Detect Evolved Android Malware}. In
  \bibinfo{booktitle}{\emph{CCS 2020}} (Virtual Event, USA).
  \bibinfo{publisher}{Association for Computing Machinery},
  \bibinfo{address}{New York, USA}, \bibinfo{pages}{757–770}.
\newblock


\bibitem[\protect\citeauthoryear{{\v{Z}}liobait{\.{e}}, Pechenizkiy, and
  Gama}{{\v{Z}}liobait{\.{e}} et~al\mbox{.}}{2016}]%
        {zliobaite2016}
\bibfield{author}{\bibinfo{person}{Indr{\.{e}} {\v{Z}}liobait{\.{e}}},
  \bibinfo{person}{Mykola Pechenizkiy}, {and} \bibinfo{person}{Jo{\~a}o Gama}.}
  \bibinfo{year}{2016}\natexlab{}.
\newblock \bibinfo{booktitle}{\emph{An Overview of Concept Drift
  Applications}}.
\newblock \bibinfo{publisher}{Springer International Publishing},
  \bibinfo{address}{Cham}, \bibinfo{pages}{91--114}.
\newblock


\end{thebibliography}

\appendix

\section{Experimental Results on the VirusShare Dataset} \label{sec:append-exp-virusshare}

\begin{figure*}[t!]
	\centering
	\includegraphics[width=0.85\textwidth]{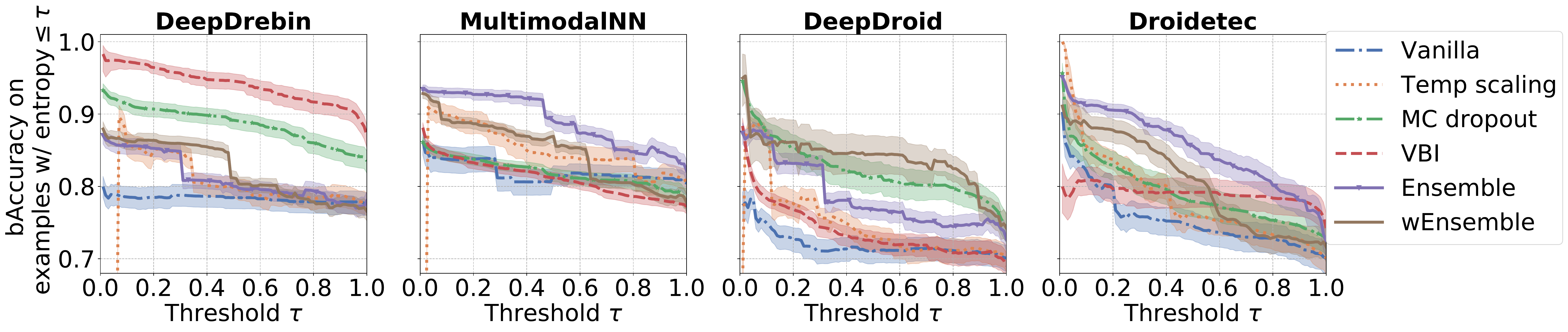}
	\captionof{figure}{The balanced accuracy on the VirusShare dataset after excluding the examples for which the detector has high uncertainties (i.e., the examples for which the predictive entropy is above a pre-determined threshold $\tau$).
	For each curve, a shadow region is obtained from the 95\% confidence interval using a bootstrapping with $10^3$ repetitions and sampling size being the number of examples in the VirusShare dataset.
	}
	\label{fig:oos-baccuracy}
\end{figure*}

Figure \ref{fig:oos-baccuracy} plots the balanced accuracy on the VirusShare dataset with decision referral. We observe that Figure \ref{fig:oos-baccuracy} exhibit the trends that are similar to what are exhibited by Figure \ref{subfig:oos-02}, except for Temp scaling on DeepDrebin and MultimodalNN.
This is because the model predicts benign examples accurately, but do not predict malicious examples accurately.

\section{Experimental Results on the Androozoo Dataset} \label{sec:append-exp-androzoo}

\begin{figure*}[t!]
	\centering
	\includegraphics[width=0.82\textwidth]{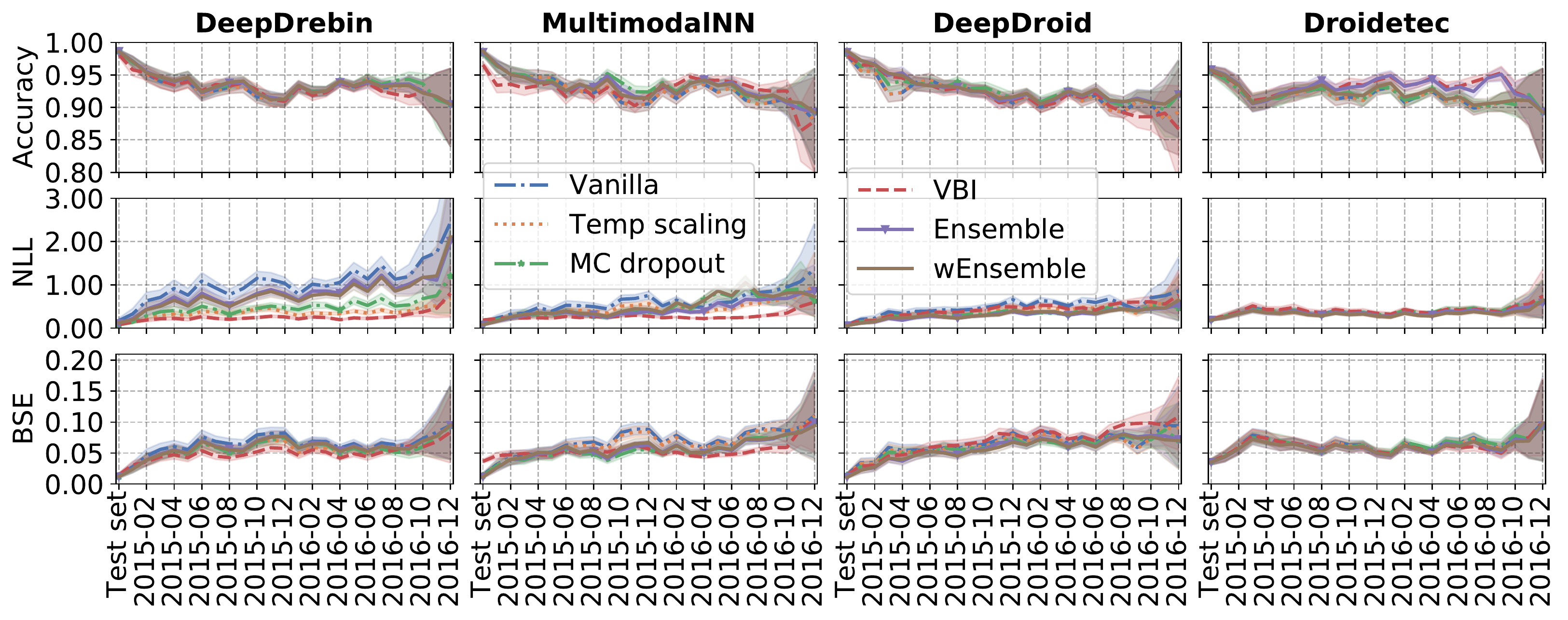}
	\caption{Illustration of accuracy, NLL, BSE under temporal covariate shift on the Androzoo dataset. A shadow region is obtained from the 95\% confidence interval using a bootstrapping with $10^3$ repetitions and sampling size being the number of APKs in per month (cf. Figure \ref{fig:androzoo-dataset}).  
		}
	\label{fig:androzoo-acc-nll-brier}
\end{figure*}

Figure \ref{fig:androzoo-acc-nll-brier} plots the Accuracy, NLL and BSE of malware detectors under temporal covariate shifts. 
We observe that the Accuracy, NLL, and BSE are smaller than their balanced counterparts plotted in Figure \ref{fig:androzoo-bacc-bnll-bbrier}, owing to the data imbalance exhibited by the Androzoo dataset, despite that they all exhibit a similar trend. 

\ignore{
\begin{figure*}[!htbp]
	\centering
	\includegraphics[width=0.8\textwidth]{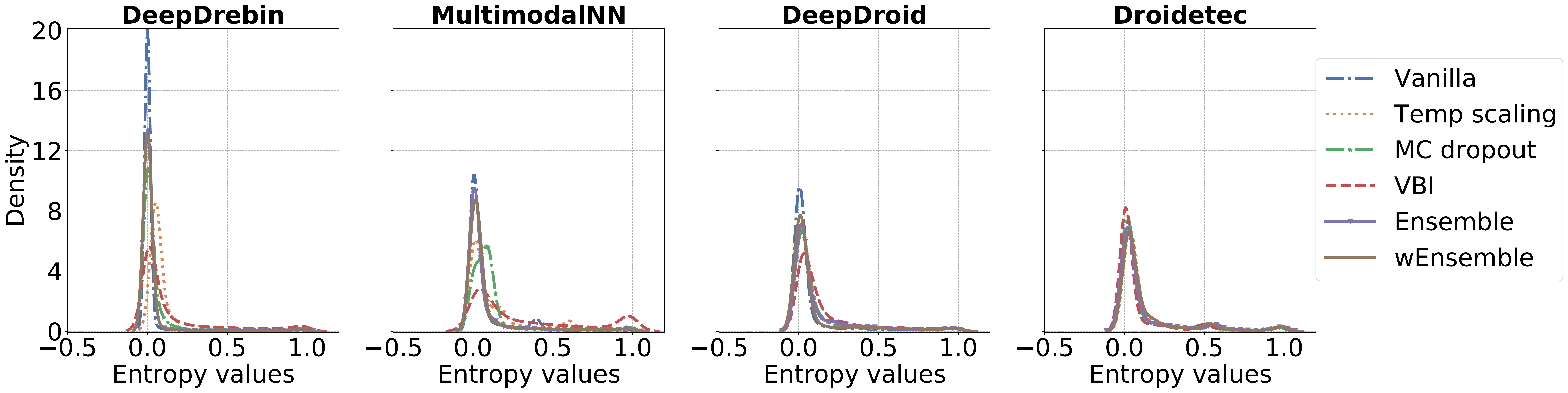}
	\caption{Sample density of predictive entropy associated with the Androzoo dataset.
	}
	\label{fig:androzoo-stat-entropy}
\end{figure*}
}

\end{document}